\documentclass[conference]{IEEEtran}
\usepackage{graphicx}
\usepackage{amsmath,amssymb,dsfont,bbm,epstopdf,pgfplots,mathtools,enumitem,mathrsfs, bbm, algpseudocode , graphicx, dblfloatfix} 
\usepackage[normalem]{ulem}
\usepackage{color}
\usepackage{cite}
\usepackage{graphics}
\usepackage{tikz}
\usepackage{pgfplots}
\usepackage{subcaption}
\usepackage[left=0.6in,right=0.6in,top=0.71in, bottom = 0.95in]{geometry}
\usetikzlibrary{shapes, arrows, decorations.markings, arrows.meta}
\usetikzlibrary{patterns} 
\allowdisplaybreaks
\graphicspath{{.}{./Figures/}} 
\allowdisplaybreaks[4] 

\newtheorem{lemma}{Lemma}

\newtheorem{theorem}{Theorem}

\newcommand{\Tu}{{\mathcal K_{\text{U}}} }

\renewcommand{\Pr}{{\mathbb{P}}}

\renewcommand{\i}{{\iota}}

\newcommand{\vect}[1]{\boldsymbol{#1}}

\definecolor{ForestGreen}{rgb}{0.0, 0.5, 0.0}

\renewcommand{\P}{\mathsf{P}}

\renewcommand{\P}{\mathsf{P}}

\newcommand{\U}{\mathsf{U}}
\newcommand{\e}{\mathsf{e}}

\newcommand{\bu}{\beta_{\U}}
\newcommand{\bef}{\beta_{\e}}

\newcommand{\xkef}{\vect X_{b}^{(\e,2)}}
\newcommand{\xkes}{\vect X_{b}^{(\e,1)}}
\newcommand{\xku}{\vect X_{b}^{(\U)}}
\newcommand{\Nu}{n_{\U}}
\newcommand{\Ne}{n_{\e}}

\newcommand{\MkS}{M^{(\e)}}
\newcommand{\syv}{ \sigma_{3}^2}
\newcommand{\sy}{ \sigma^2}
\newcommand{\vnorm}{\beta_{\text{v}}}

\newcommand{\TaU}{\mathcal T^{(\U)}}

\newcommand{\syx}{\sigma_2^2}
\newcommand{\ru}{\rho_{\U}}
\newcommand{\bku}{B_{\text{dt}}}
\newcommand{\bkut}{B_{\text{dc}}}

\newcommand{\Pb}{ \P}
\newcommand{\rd}{\rho_{\text{det}}}
\newcommand{\rdz}{\rho_{\text{det},0}}
\newcommand{\rdo}{\rho_{\text{det},1}}

\newcommand{\mb}{\mathcal B_{\text{arrival}}}
\newcommand{\mbt}{ \mathcal B_{\text{sent}}}
\newcommand{\mbd}{ \mathcal B_{\text{detect}}}
\newcommand{\mbc}{ \mathcal B_{\text{decode}}}

\newcommand{\abst}{ A_{b,\text{sent}}}
\newcommand{\abd}{ A_{b,\text{detection}}}
\newcommand{\abc}{ A_{b,\text{decode}}}
\begin{document}
\title{Joint Coding of eMBB and URLLC in Vehicle-to-Everything (V2X) Communications }
\author{\IEEEauthorblockN{Homa Nikbakht$^{1}$, Eric Ruzomberka$^{1}$, Mich\`ele Wigger$^2$, Shlomo Shamai (Shitz)$^3$, and H.~Vincent Poor$^1$}
	\IEEEauthorblockA{$^1$Princeton University,   $\quad ^2$LTCI,   T$\acute{\mbox{e}}$l$\acute{\mbox{e}}$com Paris, IP Paris,  $\quad ^3$Technion,  \\
		\{homa, er6214, poor\}@princeton.edu, michele.wigger@telecom-paris.fr,  \\ sshlomo@ee.technion.ac.il}}
\maketitle

 \begin{abstract}
A point-to-point communication is considered where a roadside unite (RSU) wishes to simultaneously send messages of enhanced mobile broadband (eMBB) and ultra-reliable low-latency communication (URLLC) services to a vehicle. The eMBB message arrives at the beginning of a block and its transmission lasts over the entire block. During each eMBB transmission block, random arrivals of URLLC messages are assumed. To improve the reliability of the URLLC transmissions, the RSU reinforces their transmissions by mitigating the interference of eMBB transmission by means of dirty paper coding (DPC).   In the proposed coding scheme, the eMBB messages are decoded based on two approaches: treating interference as noise, and successive interference cancellation. Rigorous bounds are derived for the error probabilities of eMBB and URLLC transmissions achieved by our scheme. Numerical results illustrate that they are lower than bounds for standard time-sharing.
\end{abstract}

\section{Introduction}
Enhanced mobile broadband (eMBB) and ultra-reliable low-latency communication (URLLC) services enabled by  5G new radio (NR) are considered as key enablers of the vehicle-to-everything (V2X) technology \cite{Parkvall2017, Abood2023, Noor2022, Anand2020, HomaEntropy2022, Popovski2018}. Particularly,  eMBB services aim to provide high data rate for content delivery and therefore improve the quality of experience (QoE) of in-vehicle entertainment applications. URLLC services, however, are key to guarantee the delivery of critical road safety information and thus enable fully autonomous driving of connected vehicles \cite{YChen2020, Ganesan2019}.

 Coexistence of eMBB and URLLC services in V2X communications  has been studied in the literature \cite{Chen2020, Yin2021, Song2019}. In \cite{Chen2020}, a novel URLLC and eMBB coexistence mechanism for the cellular V2X framework is proposed where at the begining of the transmission interval  eMBB users are associated with a V2X base station, whereas, URLLC users are allowed to puncture the eMBB transmissions upon arrival. The work in \cite{Yin2021} formulates an optimization problem for joint scheduling of punctured eMBB and URLLC traffic to  maximize the  aggregate utility of the eMBB users subject to latency constraints for the URLLC users. Related to this work is \cite{Song2019}, where   resources are allocated jointly between eMBB and URLLC messages  for a one-way highway vehicular network in which a vehicle receives an eMBB message from the nearest roadside unit (RSU) and URLLC messages from the nearest vehicle. During each eMBB transmission interval, random arrivals of URLLC messages are assumed. The eMBB time slot is thus divided into mini-slots and the newly arrived URLLC messages are immediately scheduled in the next mini-slot by puncturing the on-going eMBB transmissions. To guarantee the reliability of the URLLC transmission, guard zones are deployed around the vehicle and the eMBB transmissions are not allowed inside such zones. 

In this work, the RSU wishes to transmit both eMBB and URLLC messages to a vehicle. 
The eMBB message arrives at the beginning of a block and  its transmission lasts over the entire block. The eMBB blocklength is again divided into mini-slots and URLLC messages  arrive randomly at the beginning of these mini-slots. Specifically, at the beginning of each of these mini-slots a URLLC message arrives with probability $\rho \in [0,1]$ and the  
RSU simultaneously sends the eMBB message as well as the newly arrived URLLC message over this mini-slot. With probability $1-\rho$  no URLLC message arrives at the beginning of the mini-slot and the RSU only sends the eMBB message. In our work, we do not use guard zones, but instead the RSU reinforces transmission of URLLC messages by mitigating the interference of eMBB transmission by means of  dirty paper coding \cite{Costa1983, Scarlett2015, Caire2003}. After each mini-slot, the receiving vehicle attempts to decode a URLLC message, and after the  entire transmission interval  it decodes the eMBB message. Given that the URLLC transmissions interfere with the transmission of eMBB, we employ two different eMBB decoding approaches.  The first approach, known as \emph{treating interference as noise (TIN)}, is to treat the URLLC interference as noise.  The second approach, known as \emph{successive interference cancellation (SIC)},  is to first subtract the decoded   URLLC message  and then decode the eMBB message based on the received signal.  Rigorous bounds are derived for achievable error probabilities of eMBB (in both approaches) and URLLC transmissions.
Numerical results illustrate that our proposed scheme significantly outperforms the standard time-sharing scheme. 

\section{Problem Setup}\label{sec:DescriptionOfTheProblem}
Consider a point-to-point setup with one RSU (transmitter) and one vehicle (receiver) communicating over a $\Ne$ uses of an AWGN channel. The  transmitter (Tx)  sends a single, so called \emph{eMBB}-type message $\MkS$, over the entire blocklength $\Ne$, where $\MkS$  is uniformly distributed over a given set $\mathcal{M}^{(\e)} := \{1, \ldots, L_{\e}\}$. Message $\MkS$ is thus available at the Tx at time $t=1$ (and remains until time $\Ne$). Additionally, prior to each channel use in

\begin{IEEEeqnarray}{rCl}
\TaU &: =& \{1, 1+ \Nu,  1+2\Nu,\ldots, 1 + \left(\eta-1 \right)\Nu   \},
\end{IEEEeqnarray}

where 
\begin{equation}
\eta := \left \lfloor \frac{\Ne}{\Nu}\right \rfloor,
\end{equation}
 the Tx generates with probability $\rho$ an additional, so called, \emph{URLLC}-type message that it wishes to convey to the Rx. With probability $1-\rho$ no URLLC-type message is generated. 
For each $b\in [\eta]$, if a URLLC message is generated at time $t=(b-1)\Nu+1$, then we set $A_b=1$, and otherwise we set $A_b=0$. Denote  the time-instances from $(b-1)\cdot \Nu +1$ to $b\cdot \Nu$ by block~$b$. If in block $b$ a message is generated we denote it by $M_{b}^{(\U)}$ and  assume that it is  uniformly distributed over the set $\mathcal{M}^{(\U)}:= \{1, \ldots, L_{\U}\}$.

%
%

During block~$b$, the  Tx computes its  inputs  as:
\begin{IEEEeqnarray}{rCl}
X_{t} &= &\begin{cases}f^{(\U)}_t \left( M_b^{(\U)}, \MkS \right), & \text{if } A_b = 1, \\
f^{(\e)}_t \big( \MkS\big), &  \text{if } A_b = 0,
\end{cases} 
\end{IEEEeqnarray}
for  $t=(b-1)\cdot \Nu+1,\ldots, b\cdot \Nu$ and some encoding functions $f^{(\U)}_t$ and  $f_t^{(\e)}$ on appropriate domains. After the last URLLC block, i.e. at times $t=\eta \Nu +1, \ldots, \Ne$, the Tx produces the inputs 
\begin{IEEEeqnarray}{rCl}
X_{t} &= &
f^{(\e)}_t \big( \MkS\big), 
  \quad t= \eta \Nu +1, \ldots, \Ne. \IEEEeqnarraynumspace
\end{IEEEeqnarray}

The sequence of channel inputs $X_1,\ldots, X_{\Ne}$ has to satisfy 
the average block-power constraint
\begin{equation}\label{eq:power}
\frac{1}{\Ne} \sum_{t=1}^{\Ne} X_{t}^2
\leq \P, 
\qquad \textnormal{almost surely.}
\end{equation}



The input-output relation of the network is  described as
\begin{equation}\label{Eqn:Channel}
{Y}_{t} = h {X}_{t}+ {Z}_{t},
\end{equation}
where $\{Z_{t}\}$ are independent and identically distributed (i.i.d.) standard Gaussian for all  $t$ and independent of all messages; $h> 0$ is the fixed channel coefficient between the Tx and Rx.

After each URLLC block $b$ the receiver (Rx) decodes the transmitted URLLC message $M_b^{(\U)}$ if $A_b=1$. Moreover, at the end of the entire $\Ne$ channel uses it decodes the eMBB message $M^{(\e)}$. 
Thus, if $A_b=1$ it produces 
\begin{equation}
	 \hat M_b^{(\U)}={g^{(\Nu)}}\big( Y_{(b-1)\Nu +1}, \ldots, Y_{b\Nu} \big), 
	\end{equation} 
for some decoding function $g^{(\Nu)}$ on appropriate domains. Otherwise, it sets $  \hat M_b^{(\U)} = 0$. 
We define the average error probability for each message $M_b^{(\U)}$ as:
\begin{IEEEeqnarray}{rCl}
\epsilon^{(\U)}_b &:=&\rho \Pr\left[  \hat M_b^{(\U)} \neq M_b^{(\U)} \Big| A_b=1 \right] \notag \\
&&+(1-\rho) \Pr\left[  \hat M_b^{(\U)} \neq 0 \Big| A_b=0 \right] .\IEEEeqnarraynumspace
\end{IEEEeqnarray}
At the end of the  $\Ne$ channel uses,  
the Rx decodes its desired eMBB message as:
\begin{equation}\label{mhats}
\hat{{M}}^{(\e)}={\psi^{(\Ne)}}\left ( \vect Y^{\Ne} \right ),
\end{equation}
 where $\vect Y^{\Ne}: = (Y_1, \ldots, Y_{\Ne})$ and  $\psi^{(\Ne)}$ is a decoding function on appropriate domains. We define the average error probability for  message $M^{(\e)}$ as 
\begin{equation}
\epsilon^{(\e)} := \Pr\left [ \hat M^{(\e)} \neq M^{(\e)}\right].
\end{equation}

The goal is to propose a coding scheme that simultaneously has small error probabilities $\epsilon^{(\U)}_b$ and $\epsilon^{(\e)}$.

\begin{figure}[t]
  \centering
  \footnotesize
 \begin{tikzpicture}[scale=1.6, >=stealth]
\centering
\tikzstyle{every node}=[draw,shape=circle, node distance=0cm];
\foreach \i in {0,1,2,3,8,9,10,11}{
\draw [draw = none, fill = red]  (0+\i*0.25,0.05) rectangle (0.25+0.25*\i,0.25);
}
\foreach \i in {4,5,6,7,12,13,14,15,16,17,18}{
\draw [draw = none, fill = blue!60]  (0+\i*0.25,0.05) rectangle (0.25+0.25*\i,0.25);
}
\foreach \i in {0,1,2,...,18}{
\draw [draw = none, fill = blue!60]  (0+\i*0.25,-0.15) rectangle (0.25+0.25*\i,0.05);
\draw  [very thick] (0+\i*0.25,-0.15) rectangle (0.25+0.25*\i,0.25);
}

\foreach \i in {0,4,8,12,16,19}{
\draw[dashed] (\i*0.25, 0.25)--(\i*0.25, 0.6);
}
\draw [<->] (0,0.6)--(4.75,0.6);
\node [draw = none] at (2.5,0.7) {$\Ne$};
\node [draw = none] at (0.5,0.4) {$\Nu$};
\node [draw = none] at (0.5+1,0.4) {$\Nu$};
\node [draw = none] at (0.5+2,0.4) {$\Nu$};
\node [draw = none] at (0.5+3,0.4) {$\Nu$};
\node [draw = none] at (0.5+3.88,0.4) {$\Ne - \eta\Nu$};
\node [draw = none] at (0.5,-0.3) {$ \vect X_{1}^{(\U)}\hspace{-0.15cm} + \vect X_{1}^{(\e,2)}$};
\node [draw = none] at (0.5+1,-0.3) {$  \vect X_{2}^{(\e,1)}$};
\node [draw = none] at (0.5+2,-0.3) {$ \vect X_{3}^{(\U)} \hspace{-0.15cm}+\hspace{-0.05cm} \vect X_{3}^{(\e,2)}$};
\node [draw = none] at (0.5+3,-0.3) {$  \vect X_{4}^{(\e,1)}$};
\node [draw = none] at (0.5+3.9,-0.3) {$  \vect X_{5}^{(\e,1)}$};

\end{tikzpicture}
\vspace*{-5ex}

  \caption{Example of the coding scheme with $\eta = 4$ and $\mbt = \{1,3\}$.}
  \label{fig1}
  \vspace*{-2ex}
\end{figure}
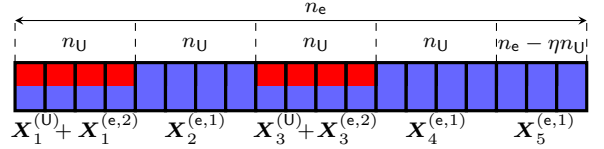~~%

\section{Joint Transmission of URLLC and eMBB Messages}\label{coding}
\subsection{Construction of Codebooks}
Define \begin{IEEEeqnarray}{rCl} \label{eq:mb}
\mb: =\{b \in [\eta]: A_b = 1\}.
\end{IEEEeqnarray}
Choose $\bu$ and $\bef \in [0,1]$ such that:
\begin{equation} \label{eq:10}
\bu + \bef = 1.
\end{equation}
Fix a value of $\alpha  \in [0,1]$. For each block $b\in [\eta]$, for each $j \in [ L_v]$ and each  realization  $m \in [ L_{\U}]$, generate  codewords $\vect V_b(m,j)$    by picking them uniformly over a centered $\Nu$-dimensional  sphere of radius $\sqrt{\Nu\vnorm  \P}$ independently of each other and  of all other codewords, for
\begin{IEEEeqnarray}{rCl}
 \vnorm : = \bu + \alpha^2 \bef.
\end{IEEEeqnarray}

For each $\ell \in [L_{\e}]$ randomly draw a  codeword $\xkef(\ell)$ uniformly distributed  on the centered $\Nu$-dimensional sphere of radius $\sqrt{\Nu \bef \P}$ and a codeword $\xkes(\ell)$ uniformly distributed  on the centered $\Nu$-dimensional sphere of radius  $\sqrt{\Nu \Pb}$. All codewords are chosen independently of each other.

\subsection{Encoding}

\subsubsection{Encoding at Blocks~$b \in \mb$} 
In each block~$b \in \mb$, the Tx has both an eMBB and an URLLC message to send.  It first picks the codeword $\xkef(\MkS)$ and then 
  employs DPC to encode $M^{(\U)}_b$  while precanceling the interference of its own eMBB codeword  $\xkef(\MkS)$. 
Specifically, it chooses an index $j$ such that the 
sequence 
\begin{equation} \label{eq:x21}
\xku :  = \vect V_b (M^{(\U)}_b,j )- \alpha  \xkef
\end{equation}
 lies in the set
\begin{IEEEeqnarray}{rCl}\label{eq:di}
\mathcal D_b := \left \{ \vect x_{b}^{(\U)}: \Nu \bu \P - \delta_{b} \le \left\|\vect x_{b}^{(\U)}\right\|^2 \le \Nu \bu \P \right \} \IEEEeqnarraynumspace
\end{IEEEeqnarray}
for a given $\delta_b> 0$. 
If multiple  such codewords  exist, the index $j^\star$ is chosen at random from this set, and the Tx sends: 
\begin{equation}
\vect{X}_b= \xku + \xkef.
\end{equation}We also set $\abst=1$.

 If
 no appropriate codeword exists, the Tx  discards the arrived URLLC message by setting $\abst=0$ and sends only the eMBB message
\begin{equation}
\vect{X}_b=\xkes(\MkS)
\end{equation}
 over this block. 


Define 
\begin{IEEEeqnarray}{rCl} \label{eq:mbt}
\mbt := \{b \in \mb:  \abst = 1\},
\end{IEEEeqnarray}
where $\mbt \subseteq \mb$ and represents the set of blocks in which an URLLC message is sent. See Figure~\ref{fig1}.

\subsubsection{Encoding at Blocks~$b \in [\eta] \backslash \mb$ and in Block $\eta+1$ when $\Ne > \eta \Nu$} 
In each Block~$b \in [\eta] \backslash \mb$, the Tx sends only eMBB message $M^{(\e)}$: 
\begin{equation}
\vect{X}_b=\vect X_{b,1}^{(\e)}(M^{(\e)}).
\end{equation}

%

Over Block~$b$, the Tx  thus transmits
\begin{IEEEeqnarray}{rCl}
 \vect X_b =\begin{cases} \xku+ \xkef & \text{if} \; b \in \mbt, \\
\xkes & \text{o.w.}
\end{cases}
\end{IEEEeqnarray}


\subsection{Decoding} 
After each block~$b \in [\eta]$, the Rx 
  attempts to decode a URLLC message, and after the  entire block of $\Ne$ channel uses it decodes the transmitted  eMBB message. Given that the URLLC transmissions interfere with the transmission of eMBB, the Rx envisions  two different approaches to decode the eMBB message. The first approach, termed  \emph{TIN approach},  is to treat the URLLC interference as noise.  The second approach, termed \emph{SIC approach}, is to first subtract the decoded   URLLC message  and then decode the eMBB message based on the received signal. 
\subsubsection{Decoding of URLLC Messages} At the end of each block~$b \in [\eta]$, the Rx observes the following channel outputs $\vect Y_b: = \{Y_{(b-1)\Nu + 1}, \ldots, Y_{b \Nu} \}$: 

\begin{IEEEeqnarray}{rCl}
\vect Y_b = \begin{cases} h\xku + h\xkef + \vect Z_b \quad & \text{if}\; b \in \mbt  \\
h \xkes +  \vect Z_b \quad & \text{o.w.}  \end{cases}
\end{IEEEeqnarray}
with $\vect Z_b \sim \mathcal N(0, I_{\Nu})$. 
Define the information density metric between $\vect y_b$ and $\vect v_b$ by:
\begin{equation} \label{eq:ibU}
i^{(\U)}_b  (\vect v_b; \vect y_b ) := \ln \frac{f_{\vect Y_b| \vect V_b} (\vect y_b| \vect v_b)}{f_{\vect Y_b}(\vect y_b)}. 
\end{equation}

After observing $\vect Y_b$,  the Rx  chooses  the pair
\begin{equation}
(m',j') =\text{arg} \max_{ m, j}  i^{(\U)}_b  (\vect v_b(m,j); \vect Y_b ) .
\end{equation}
If for this pair 
\begin{equation}
 i^{(\U)}_b  (\vect v_b(m',j'); \vect Y_b ) > \gamma^{(\U)}
 \end{equation}
 where $\gamma^{(\U)}$ is a threshold over which we optimize, the Rx chooses $(\hat M_b^{(\U)},\hat j)= (m',j')$ and sets $\abd = 1$. Otherwise the receiver declares that no URLLC message has been sent and indicates it by setting $\hat M_b^{(\U)}=0$ and $\abd = 0$.

Define
\begin{IEEEeqnarray}{rCl} \label{eq:mbd}
\mbd := \{b \in [\eta]: \abd = 1\}
\end{IEEEeqnarray}
that is the set of blocks in which an URLLC message is detected. A detection error happens if 
$\mbd \neq \mbt$.

In each block~$b \in \mbd$, set $\abc = 1$ if $(\hat M_b^{(\U)}, \hat j) = (M_b^{(\U)} , j)$, otherwise set $\abc = 0$. Define 
\begin{IEEEeqnarray}{rCl} \label{eq:mbc}
\mbc : = \{b \in \mbd: \abc = 1\}\IEEEeqnarraynumspace
\end{IEEEeqnarray}
that is the set of blocks in which an URLLC message is decoded correctly. 

\subsubsection{Decoding the eMBB Message under the TIN approach} \label{sec:eMBBTIN}
To decode its desired eMBB message under this approach, the Rx treats URLLC transmissions as noise. Therefore,
 the decoding of the eMBB message depends on the detection of URLLC messages sent over the $\eta$ blocks.

Let $\bku$ be the realization of the set $\mbd$ defined in \eqref{eq:mbd}.
Given $\bku$, the Rx decodes its desired eMBB message based on the outputs of the entire $\Ne$ channel uses by  looking for an index $m$ such that its corresponding codewords $\left \{ \{\vect x_{b}^{(\e,1)}(m)\}_{b \notin  \bku },  \{\vect x_{b}^{(\e,2)}(m)\}_{b \in \bku } \right \}$ maximize 
\begin{IEEEeqnarray}{rCl}
\lefteqn{i^{(\e)}_{\text{TIN}} \left ( \{\vect x_{b}^{(\e,1)}\}_{b \notin  \bku },  \{\vect x_{b}^{(\e,2)}\}_{b \in \bku };  \vect y^{\Ne}| \mbd = \bku \right)}\notag \\ &&
: = \ln \hspace{-0.15cm}\prod_{b\notin  \bku  }\hspace{-0.15cm} \frac{f_{\vect Y_b| \xkes} (\vect y_b| \vect x_{b,1}^{(\e)})}{f_{\vect Y_b}(\vect y_b)} +  \ln \hspace{-0.15cm}\prod_{b\in  \bku } \hspace{-0.15cm} \frac{f_{\vect Y_b| \xkef} (\vect y_b| \vect x_{b,2}^{(\e)})}{f_{\vect Y_b}(\vect y_b)} \IEEEeqnarraynumspace
\end{IEEEeqnarray}
among all codewords
 $ \{ \{\vect x_{b}^{(\e,1)}(m')\}_{b \notin  \bku },  \{\vect x_{b}^{(\e,2)}(m')\}_{b \in \bku } \}$.

\subsubsection{Decoding the eMBB Message under the SIC approach}\label{sec:eMBBSIC}
Under this approach, before decoding the desired eMBB message, the Rx mitigates the interference of the correctly decoded URLLC messages from its observed output signal. Therefore, the decoding of the eMBB message depends not only on the  detection of the sent URLLC messages but also on the  decoding of such messages.   

For each Block~$b \in \mbd$, we define $\abc = 1$ if $(\hat M_b^{(\U)}, \hat j) = (M_b^{(\U)} , j)$, otherwise set $\abc = 0$. Define the set of blocks in which an URLLC message is decoded correctly:
\begin{IEEEeqnarray}{rCl} \label{eq:mbc}
\mbc : = \{b \in \mbd: \abc = 1\}. \IEEEeqnarraynumspace
\end{IEEEeqnarray}

Let  $\bku$ be a realization of the set $\mbd$ and $\bkut$ be a realization of the set $\mbc$. 
After observing the channel outputs of the entire $\Ne$ channel uses, the Rx decodes its desired eMBB message  by  looking for an index $m$ such that its corresponding codewords $\left \{ \{\vect x_{b}^{(\e,1)}(m)\}_{b \notin \bku},  \{\vect x_{b}^{(\e,2)}(m)\}_{b \in \bku } \right \}$ maximize 
\begin{IEEEeqnarray}{rCl}
\lefteqn{i^{(\e)}_{\text{SIC}} \Big( \{\vect x_{b}^{(\e,1)}\}_{b \notin \bku  },  \{\vect x_{b}^{(\e,2)}\}_{b \in \bku }; \vect y^{\Ne}| \bku , \bkut, \{\vect V_b\}_{b \in \bkut}\Big)} \notag \\ &&
: = \ln \hspace{-0.15cm}\prod_{b\notin \bku  }\hspace{-0.15cm} \frac{f_{\vect Y_b| \xkes} (\vect y_b| \vect x_{b}^{(\e,1)})}{f_{\vect Y_b}(\vect y_b)} +  \ln \hspace{-0.15cm}\prod_{b\in  \bku \backslash \bkut } \hspace{-0.15cm} \frac{f_{\vect Y_b| \xkef} (\vect y_b| \vect x_{b}^{(\e,2)})}{f_{\vect Y_b}(\vect y_b)} \notag \\
&&+  \ln \hspace{0cm}\prod_{b\in  \bkut } \hspace{0cm} \frac{f_{\vect Y_b| \xkef, \vect V_b} (\vect y_b| \vect x_{b}^{(\e,2)}, \vect v_b)}{f_{\vect Y_b| \vect V_b}(\vect y_b| \vect v_b)}
\end{IEEEeqnarray}
\vspace{0.2cm}
among all codewords $\{ \{\vect x_{b}^{(\e,1)}(m')\}_{b \notin \bku  },  \{\vect x_{b}^{(\e,2)}(m')\}_{b \in \bku } \}$.

\begin{figure*}[b!] 
\hrule
\begin{subequations}\label{eq:37}
\begin{IEEEeqnarray}{rCl}
J_{\e} & :=&\left( \frac{\pi 2^{\frac{\Nu+1}{2}}e^{\frac{-h^2 \vnorm \Pb \Nu}{2}} \sqrt{\vnorm \bef}}{9h^2 (1- \alpha)^{\Nu -1} (\vnorm + (1- \alpha)^2 \bef )} \right)^k  \cdot \left ( \frac{\sqrt{8 (1 + 2 h^2\Pb)}}{27\sqrt{\pi} (1+ h^2 \Pb)} \right)^{\eta-k}\\
\tilde J_{\e} & :=& \left ( \frac{\pi 2^{\frac{\Nu+1}{2}}e^{\frac{-h^2 \vnorm \Pb \Nu}{2}} \sqrt{\vnorm \bef}}{9h^2 (1- \alpha)^{\Nu -1} (\vnorm + (1- \alpha)^2 \bef )} \right)^{ k-\tilde k} \cdot   \left ( \frac{\sqrt{8 (1 + 2 h^2\Pb)}}{27\sqrt{\pi} (1+ h^2 \Pb)} \right)^{\eta-k } \cdot \left ( \frac{\sqrt{8 (1 + 2 h^2(1-\alpha)^2 \bef \Pb)}}{27\sqrt{\pi} (1+ h^2(1-\alpha)^2\bef \Pb)} \right )^{\tilde k} \\
\zeta &: =& \frac{1}{\sqrt{\pi }}\frac{\Gamma(\frac{\Nu}{2})}{\Gamma (\frac{\Nu-1}{2})} \left(\kappa_{\frac{\Nu-3}{2}} \left (\alpha\sqrt{{\bef}/{\vnorm}} + \delta_{b}/(2 \alpha \Nu \Pb \sqrt{\vnorm\bef})  \right )  -  \kappa_{\frac{\Nu-3}{2}} \left (\alpha\sqrt{{\bef}/{\vnorm}}\right ) \right) \label{eq:zeta}\\
 \mu_{\U}& := & \frac{2\sy \syv}{h^2(\sy-\syv)} \left(\frac{\Nu}{2}\ln \frac{\sy}{\syv} - \gamma^{(\U)} + \ln J_{\U} \right)  + \frac{\syv}{\sy - \syv} \left ( \Nu \Pb (\sqrt{\bu} - \sqrt{\bef})^2 - \delta_b\right) -  \frac{\sy \Nu \bef \Pb (1-\alpha)^2}{\sy - \syv}  \label{eq:bargu}\\
\tilde  \mu_{\U} & := & \frac{2\sy \syv}{h^2(\sy-\syv)} \left(\frac{\Nu}{2}\ln \frac{\sy}{\syv} - \gamma^{(\U)} + \ln \tilde J_{\U} \right)  + \frac{\syv}{\sy - \syv} \left ( \Nu \Pb (\sqrt{\bu} +\sqrt{\bef})^2 \right)  -  \frac{\sy \Nu \bef \Pb (1-\alpha)^2}{\sy - \syv} \\
\mu&:=&  \frac{\Ne}{2} \ln \sy - \frac{k\Nu}{2} \ln \syx - \frac{\eta-k}{2\sy} \Nu \Pb  +   \frac{k}{2\syx}\vnorm \Nu \Pb - \frac{k}{2\sy}\left (\sqrt{\vnorm} + (1- \alpha)\sqrt{\bef} \right)^2\Nu \Pb -\gamma^{(\e)} + \ln J_{\e}  \label{eq:gamma} \\
\tilde \mu&: =&  \frac{\Ne}{2} \ln \sy  + \Nu \P \left (\frac{k - \tilde k}{2}\left (   \frac{\vnorm}{\syx} - \frac{\left (\sqrt{\vnorm} + (1- \alpha)\sqrt{\bef} \right)^2 }{\sy}   -\frac{\ln {\syx}}{\Pb} \right) + \frac{\tilde k}{2\P} \ln \frac{\syv}{\sy}  - \frac{\eta-k}{2\sy}  - \frac{\tilde k(1- \alpha)^2 \bef }{2 \syv}\right)  + \ln e^{-\tilde  \gamma^{(\e)}} \tilde J_{\e}\IEEEeqnarraynumspace \\
T &:= &  \frac{(\Ne - k \Nu)(\sy-1)}{2\sy \mu} +  \frac{(\eta+1 - k)\sqrt{\Nu \Pb}}{\sy \mu} \frac{\sqrt{2} \Gamma\left (\frac{\Nu +1}{2}\right)}{\Gamma\left (\frac{\Nu}{2}\right)} + \frac{k\tau}{\mu}\frac{\sqrt{2} \Gamma\left (\frac{\Nu +1}{2}\right)}{\Gamma\left (\frac{\Nu}{2}\right)} +  \frac{k \Nu (\sy-\syx)}{2\sy \syx \mu} + (L_e -1) e^{-\gamma^{(\e)}} \label{eq:T1} \IEEEeqnarraynumspace \\
\nu & :=& \frac{\tilde k}{\tilde \mu} \left ( \frac{\sqrt{2} \Gamma\left (\frac{\Nu +1}{2}\right)}{\Gamma\left (\frac{\Nu}{2}\right)}  \left (\tau -\frac{(1-\alpha)\sqrt{\Nu \bef \Pb}}{\syv}\right ) + \Nu\left ( \frac{\sy-\syx}{2\sy \syx } - \frac{\syv-1}{2\syv}\right) \right)  + (L_e -1) \left ( \frac{\mu}{\tilde \mu} e^{-\tilde \gamma^{(\e)}}+ e^{-\tilde \gamma^{(\e)}} \right) \label{eq:T2}
\end{IEEEeqnarray}
\end{subequations}
\end{figure*}
\section{Main Results} \label{sec:main}
Define $\sy := h^2  \Pb + 1$, $\syx := h^ 2\vnorm \P+ 1$, $\syv := h^ 2(1 - \alpha)^2 \bef \Pb+ 1$ and 
\begin{subequations} \label{eq:31}
\begin{IEEEeqnarray}{rCl}
\lambda (x) & := &\frac{x}{2} + \frac{u^2}{4} - \frac{u}{2} \sqrt{x+ \frac{u^2}{4} },  \\
\tilde \lambda (x) &:=& \frac{x}{2} + \frac{u^2}{4} + \frac{u}{2} \sqrt{x + \frac{u^2}{4} }, \\
u & := &   \frac{2\sqrt{\Nu \Pb} \left ( \syv (\sqrt{\bu} + \sqrt{\bef}) + \sy \sqrt{\bef}(1- \alpha)\right)}{h (\sy - \syv)}, \IEEEeqnarraynumspace\\
\tau&: =& \frac{\sqrt{\Nu \P} \left(\sqrt{\vnorm} (\sy + \syx) + (1- \alpha) \sqrt{\bef}\syx \right)}{\sy \syx},
  \end{IEEEeqnarray}
  and for all integer values $n=1,2,\ldots$:
  \begin{IEEEeqnarray}{rCl}
\kappa_n(x) &:=& \frac{x(1-x^2)^n}{2n+1} + \frac{2n}{2n+1} \kappa_{n-1}(x)  
\end{IEEEeqnarray}
\end{subequations}
where $\kappa_0(x) := x$. 
By employing the scheme proposed in Section~\ref{coding}, we have the following theorem on the upper bounds on the URLLC and eMBB error probabilities $\epsilon^{(\U)}_b$, $\epsilon^{(\e)}_{\text{TIN}}$, and $\epsilon^{(\e)}_{\text{SIC}}$.

\begin{theorem}\label{th1}
For fixed $\bef$ , $\bu \in [0,1] $ and message set sizes $L_{\U}$ and $L_{\e}$, the average error probabilities $\epsilon^{(\U)}_b$, $\epsilon^{(\e)}_{\text{TIN}}$, and $\epsilon^{(\e)}_{\text{SIC}}$ are bounded by 
\begin{IEEEeqnarray}{rCl}
\epsilon^{(\U)}_b & \le & \rho \left (( 1-  \zeta )^{L_v}  + q + 1- q_2 \right) + (1- \rho)q_1 \label{boundu}  \IEEEeqnarraynumspace\\
\epsilon^{(\e)}_{\text{TIN}} & \le & \sum_{k = 0}^\eta \binom \eta {k} q_3^{k} (1-\ru q_2)^{\eta-k} \left (1- \Delta + T \right )\IEEEeqnarraynumspace\label{boundeTIN} \\
\epsilon^{(\e)}_{\text{SIC}} & \le &\sum_{k= 0}^\eta \binom \eta {k} q_4^{k} (1-\ru q_2)^{\eta-k}\notag \\
&&\hspace{0cm}\cdot \left(1- \Delta + \sum_{\tilde{k}=0}^{k} \binom {k} {\tilde{k}} q^{\tilde{k}} (1- q)^{k - \tilde{k}}  \left (\frac{\mu T }{\tilde \mu} -\nu \right) \right ), \IEEEeqnarraynumspace \label{boundeSIC}
\end{IEEEeqnarray}
 where  $\gamma^{(\U)}, \gamma^{(\e)},  \tilde{\gamma}^{(\e)}$ are arbitrary positive parameters, $G(\cdot, \cdot)$  denotes the regularized gamma function,  $k:=|\bku|$, $\tilde k = |\bkut|$, $\ru := \rho\left(1- (1-\zeta)^{L_v}\right)$, $q_3: = \ru q_4 + (1- \ru)q_1$, and
\begin{subequations}\label{eq:defs}
\begin{IEEEeqnarray}{rCl}
q&: = &\sqrt[L_v L_{\U}]{1-q_2}  + (L_vL_{\U} -1) e^{-\gamma^{(\U)}}, \\
q_1 &:= &1- \left  (1- e^{-\gamma^{(\U)}} \right) ^{L_vL_{\U}},\\
q_2 &:=& 1 - \left  (1- G\left(\frac{\Nu}{2}, \lambda(\mu_{\U})\right) + G\left(\frac{\Nu}{2}, \tilde \lambda(\mu_{\U})\right) \right) ^{L_vL_{\U}} \\
q_4 &:=& 1 - \left  (1- G\left(\frac{\Nu}{2}, \tilde \lambda (\tilde \mu_{\U})\right) + G\left(\frac{\Nu}{2},  \lambda (\tilde \mu_{\U})\right) \right) ^{L_vL_{\U}} \\
\Delta &:=& \frac{\ru^{k}(1-\ru)^{\eta-k} q_2^{k}  (1-q_1)^{\eta-k}}{(\ru \cdot q_3 + (1- \ru)\cdot q_1)^{k } (1-\ru \cdot q_2)^{\eta- k}} \\
J_{\U} & :=&  \frac{ \pi \sqrt{ \vnorm \bef}  2^{\frac{\Nu+1}{2}}e^{-\frac{h^2(1-\alpha)^2\bef \Pb \Nu}{2}}}{9h^2(1-\alpha) (\vnorm + (1-\alpha)^2 \bef  )}, \\
\tilde J_{\U} & :=&\frac{27 \sqrt{\pi} (1+h^2(1-\alpha)^2\bef \P)e^{\Nu h^2 \P (\vnorm + (1-\alpha)^2 \bef)}}{2(h^2(1-\alpha))^{\Nu-2}\sqrt{8 (1+2h^2(1-\alpha)^2\bef\Pb} }.\IEEEeqnarraynumspace
\end{IEEEeqnarray}
\end{subequations}
and $J_e, \tilde J_e, \zeta, \mu_{\U}, \tilde \mu_{\U},  \mu, \tilde \mu, T$ and $\nu$ are defined in \eqref{eq:37}.

\end{theorem}
\begin{IEEEproof}
See Section~\ref{sec:proofth1}.
\end{IEEEproof}
\section{Numerical Analysis}
\begin{figure}[t!]
\centering
\begin{tikzpicture}[scale=0.9]
\begin{axis}[
    xlabel={\small {$\rho$ }},
    ylabel={\small {$\epsilon_{b}^{(\U)}, \epsilon_{\text{TIN}}^{(\e)}, \epsilon_{\text{SIC}}^{(\e)}$ }},
     xlabel style={yshift=.5em},
     ylabel style={yshift=0em},
    xmin=0.2, xmax=1,
    ymin=1e-12, ymax=1,
    xtick={0.2,0.4,0.6,0.8,1},
    ytick={1e-10,1e-8,1e-6,1e-4,1e-2,1},
    yticklabel style = {font=\small,xshift=0.25ex},
    xticklabel style = {font=\small,yshift=0.25ex},
    legend pos=north east,
    ymode=log,
    legend pos=south east,
]


\addplot[ color=black, very thick,  mark=diamond, dashed] coordinates { (0.2,5.76913699757873e-10)(0.4,3.51655062345994e-7)(0.6,9.24224087764362e-5)(0.8,0.0135448415730991)(1,1)};

\addplot[ color=green, very thick] coordinates { (0.2,1e-5)(0.4,1e-5)(0.6,1e-5)(0.8,1e-5)(1,1e-5)};

\addplot[ color=blue, very thick,  mark=star] coordinates { (0.2,3.000347552198156492e-10)(0.4,6.000504102538182601e-8)(0.6,2.000451018436627799e-6)(0.8,2.000397202512178848e-5)(1,0.000199822919231279)};

\addplot[ color=red, very thick,  mark=halfcircle] coordinates { (0.2,7.41089720909776e-11)(0.4,1.000251453295273685e-8)(0.6,1.000202443326283710e-6)(0.8,0.9152793012261354e-4)(1,0.00894641132049324)};

%

\legend{\small {Time Sharing},  {$\epsilon_{b}^{(\U)}$}, {\footnotesize $\epsilon^{(\e)}_{\text{TIN}}$}, {\footnotesize $\epsilon^{(\e)}_{\text{SIC}}$}}  
\end{axis}

\vspace{-0.8cm}
\end{tikzpicture}

\caption{ Upper bounds on $\epsilon_{\text{TIN}}^{(\e)}, \epsilon_{\text{SIC}}^{(\e)}$ for $\Pb = 5, \Ne = 600$ and $\Nu = 200$  and for maximum value of $\epsilon_{b}^{(\U)}$ fixed at $10^{-5}$.}
\label{fig2}
\vspace{-0.5cm}
\end{figure}
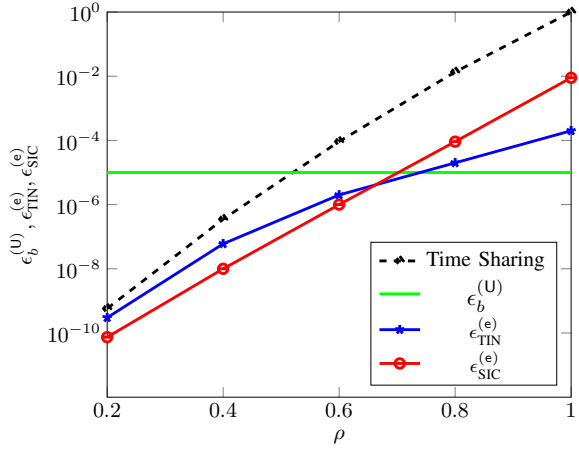

\begin{figure}[t!]
\centering
\begin{tikzpicture}[scale=0.9]
\begin{axis}[
    xlabel={\small {$\epsilon_{b}^{(\U)}$ }},
    ylabel={\small {$\epsilon_{\text{TIN}}^{(\e)},  \epsilon_{\text{SIC}}^{(\e)}$ }},
     xlabel style={yshift=.5em},
     ylabel style={yshift=0em},
    xmin=5e-6, xmax=1e-1,
    ymin=1e-12, ymax=1,
    xtick={1e-9,1e-8,1e-7,1e-6,1e-5,1e-4,1e-3,1e-2,1e-1,1},
    ytick={1e-10,1e-8,1e-6,1e-4,1e-2,1},
    yticklabel style = {font=\small,xshift=0.25ex},
    xticklabel style = {font=\small,yshift=0.25ex},
    legend pos=south east,
    ymode=log,
xmode = log,
legend columns=2, 
]


\addplot[ color=black, very thick,  dashed] coordinates { (1e-05,0.0135448415730991)(1.175194363069249e-04,0.042694161232047 )(7.230934120238305e-04,0.069082011213920)(0.004650839910841,0.108666672622638)(0.032956352334015,0.191516289054611)};

\addplot[ color=black, very thick,  mark=diamond, dashed] coordinates { (1e-05,5.76913699757873e-10)(1.175194363069249e-04,3.678313428907042e-08 )(7.230934120238305e-04,8.057194081666355e-07)(0.004650839910841,1.845531469918285e-05)(0.032956352334015,0.0005898191832318)};

\addplot[ color=blue, very thick,  mark=triangle] coordinates { (1e-05,2.000397202512178848e-5)(1.175194363069249e-04,3.644257851735579e-04 )(7.230934120238305e-04,1.597492097939694e-03)(0.004650839910841,0.005658760882520)(0.032956352334015,0.009820248651605)};

\addplot[ color=blue, very thick,  mark=star] coordinates { (1e-05,3.000347552198156492e-10)(1.175194363069249e-04, 1.530870780231997e-08)(7.230934120238305e-04,3.537520523712423e-07)(0.004650839910841,6.686960812337499e-06)(0.032956352334015,0.0002903140466673)};

\addplot[color=red, very thick,  mark=square] coordinates { (1e-05,0.9152793012261354e-4)(1.175194363069249e-04,2.926032122764477e-03 )(7.230934120238305e-04,0.01315771701962)(0.004650839910841,0.03906899890801)(0.032956352334015,0.06522675979698)};

\addplot[ color=red, very thick,  mark=otimes] coordinates { (1e-5,7.41089720909776e-11)(1.175194363069249e-04, 2.723680881461321e-09)(7.230934120238305e-04,5.178313428907042e-08)(0.004650839910841,1.956009996868224e-06)(0.032956352334015,0.0001145485767219)};

\footnotesize
\legend{ {TS, $\rho = 0.8$}, {TS, $\rho = 0.2$}, {TIN, $\rho = 0.8$},   {TIN, $\rho = 0.2$},{SIC, $\rho = 0.8$},{SIC, $\rho = 0.2$} }  
\end{axis}

\vspace{-0.8cm}
\end{tikzpicture}

\caption{ Upper bounds on $\epsilon_{\text{TIN}}^{(\e)}, \epsilon_{\text{SIC}}^{(\e)}$, $\epsilon_{b}^{(\U)}$ for $\P = 5$ and $\Nu = 20\cdot b$ and $\Ne = 3\Nu$ for values of  $b$ in $\{ 10,8,6,4,2\}$. }
\label{fig3}
\vspace{-0.5cm}
\end{figure}
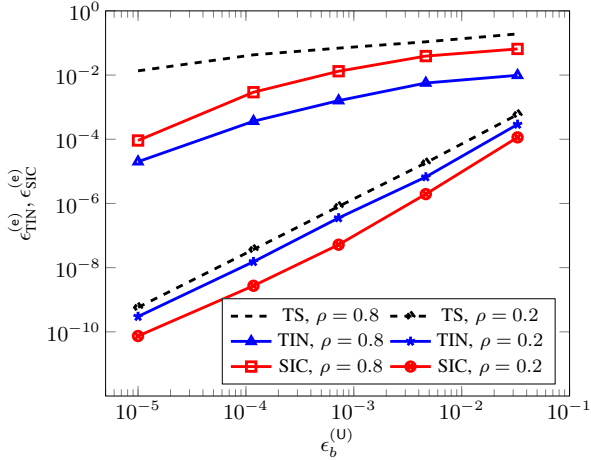

In Figure~\ref{fig2}, we numerically compare the bounds in Theorem~\ref{th1} with the time-sharing scheme where URLLC transmissions puncture the eMBB transmission upon arrival. In this figure,  we set the maximum error probability of URLLC transmission to be equal to $10^{-5}$. For each value of $\rho \in \{0.2,0.4,0.6,0.8,1\}$, we then  optimize the parameters $\alpha$, $\bef$ and $\bu$ to minimize the eMBB error probability under both TIN and SIC approaches. As can be seem from this figure, our schemes outperform the time-sharing scheme specifically for large values of $\rho$, i.e., in regions with dense URLLC arrivals.

In Figure~\ref{fig3},  we numerically compare the bounds in Theorem~\ref{th1} for $\rho = 0.2$ and $\rho = 0.8$. In this plot, $\Nu = 20 \cdot b$ and $\Ne = 3\Nu$ and the value of $b$ varies from $10$ to $2$ with step size $2$. The values of $\alpha$, $\bef$ and $\bu$ are optimized to minimize  $\epsilon_{\text{TIN}}^{(\e)}$ and $ \epsilon_{\text{SIC}}^{(\e)}$ for a given maximum $\epsilon_{b}^{(\U)}$.  As can be seen from this figure, when $\rho$ is high, the TIN scheme outperforms the SIC and the time-sharing schemes. For low values of $\rho$, however, the SIC scheme outperforms the other two schemes. The reason is that for high values of  $\rho$, more subtracted URLLC interference will be wrong which introduces error in the eMBB decoding under the SIC scheme. 

\section{Proof of Theorem~\ref{th1}} \label{sec:proofth1}
\subsection{Bounding $\epsilon_b^{(\U)}$}
Recall the definition of the sets $\mb$, $\mbt$ and $\mbd$ from \eqref{eq:mb}, \eqref{eq:mbt} and \eqref{eq:mbd}, respectively.
Given that URLLC message $M_b^{(\U)}$ arrives at the beginning of Block~$b$, i.e., $b \in \mb$,  we have the following error events: 
\begin{IEEEeqnarray}{rCl}
\mathcal E_{\U,1} &: =& \{b \notin \mbt \}  \\
\mathcal E_{\U,2} &:=& \{ b \notin \mbd \} \label{eq:error2} \\
\mathcal E_{\U,3} &: = &\left \{ \left(\hat M_b^{(\U)}, \hat j \right) \neq \left (M_b^{(\U)}, j \right) \right\}. \label{eq:error3}
\end{IEEEeqnarray}
Given that no URLLC message is sent over  Block~$b$, i.e., $b \notin \mbt$,  we have the following error event: 
\begin{IEEEeqnarray}{rCl}
\mathcal E_{\U,4} &: = & \{b \in \mbd\} .\label{eq:error4}
\end{IEEEeqnarray}
The  error probability of decoding URLLC message $M_b^{(\U)}$ of Block $b$  thus is bounded by 
\begin{IEEEeqnarray}{rCl}
\epsilon_b^{(\U)} &\le &  \Pr [b \in \mb] \Pr [\mathcal E_{\U,1} | b \in \mb] \notag \\
&& + \Pr [b \in \mb] \Pr [\mathcal E_{\U,2}  | \mathcal E_{\U,1}^c, b \in \mb] \notag \\
&& +  \Pr [b \in \mb] \Pr [\mathcal E_{\U,3}| \mathcal E_{\U,2}^c, \mathcal E_{\U,1}^c, b \in \mb ] \notag \\
&& +  \Pr [b \notin \mb] \Pr [\mathcal E_{\U,4} | b \notin \mb] .
\end{IEEEeqnarray}
\subsubsection{Analyzing $\Pr [\mathcal E_{\U,1} | b \in \mb]$} 
  From \eqref{eq:di} we notice  that $\Big(\vect V_b - \alpha  \xkef \Big)\in \mathcal D_b$ if and only if 
\begin{IEEEeqnarray}{rCl} \label{eq:25}
 \Nu \bu \Pb - \delta_{b} \le ||\vect V_b - \alpha  \xkef ||^2  \le \Nu \bu \Pb.
\end{IEEEeqnarray}
Recall that  $||\vect V_k||^2  = \Nu \vnorm \Pb$ almost surely. 
\begin{lemma}\label{lemma1}
We can prove that
\begin{IEEEeqnarray}{rCl}
\Pr [(\vect V_b - \alpha  \xkef ) \in \mathcal D_b ] = \zeta
\end{IEEEeqnarray}
where $\zeta$ is defined in \eqref{eq:zeta}.
\end{lemma}
\begin{IEEEproof}
see Appendix~\ref{App:A}.
\end{IEEEproof}

Since the $L_v$ codewords are generated independently:
\begin{IEEEeqnarray}{rCl} \label{eq:e1}
\Pr [\mathcal E_{\U,1} | b \in \mb] = \left ( 1-  \zeta \right )^{L_v}.
\end{IEEEeqnarray}

To analyze the remaining error events, we employ the following lemma. 
\begin{lemma} \label{lemma2}
For any $\gamma^{(\U)}>0$: 
\begin{IEEEeqnarray}{rCl}
\lefteqn{\Pr [i_b^{(\U)}(\vect V_b(m,j); \vect Y_b) \le \gamma^{(\U)} ] } \nonumber \\
&\le& 1-G\left(\frac{\Nu}{2}, \lambda (\mu_{\U} )\right) + G\left(\frac{\Nu}{2}, \tilde \lambda  (\mu_{\U} )\right),
\end{IEEEeqnarray}
where $G(\cdot,\cdot)$ is the regularized gamma function 
and $\lambda(\cdot)$ and $ \tilde \lambda(\cdot)$ are defined in \eqref{eq:31} and $\mu_{\U}$ is defined in \eqref{eq:37}. 
\end{lemma}
\begin{IEEEproof}
See Appendix~\ref{App:B}.
\end{IEEEproof}

\subsubsection{Analyzing $\Pr [\mathcal E_{\U,2}  | \mathcal E_{\U,1}^c, b \in \mb]$}
This error event is equivalent to the probability that for all $j \in [L_v]$ and for all $m \in [L_{\U}]$ there is no codeword $V_b(m,i)$ such that $i(\vect V_b(m,i); \vect Y_b) > \gamma^{(\U)}$. Therefore,
\begin{IEEEeqnarray}{rCl}
\lefteqn{\Pr [\mathcal E_{\U,2}  | \mathcal E_{\U,1}^c, b \in \mb]} \notag \\
&=&\left( \Pr\left [i(\vect V_b(m,j); \vect Y_b) \le \gamma^{(\U)}\right]\right)^{L_vL_{\U}} \\
&\le& \left  (1- G\left(\frac{\Nu}{2}, \lambda (\mu_{\U}) \right) + G\left(\frac{\Nu}{2}, \tilde \lambda (\mu_{\U})\right) \right) ^{L_vL_{\U}}\label{eq:e3}
\end{IEEEeqnarray}
where the last inequality holds by Lemma~\ref{lemma2}. 

\subsubsection{Analyzing $\Pr [\mathcal E_{\U,3}| \mathcal E_{\U,2}^c, \mathcal E_{\U,1}^c, b \in \mb ]$}
To evaluate this probability, we use the threshold bound for maximum-metric decoding. For any given threshold $\gamma^{(\U)} $:  
\begin{IEEEeqnarray}{rCl} \label{eq:43}
\lefteqn{\Pr [\mathcal E_{\U,3}| \mathcal E_{\U,2}^c, \mathcal E_{\U,1}^c, b \in \mb ]} \\
&\le& \Pr[i(\vect V_b(M_b^{(\U)},  j); \vect Y_b) \le \gamma^{(\U)}]  \notag \\
&&+ (L_vL_{\U} -1) \mathbb P[i(\bar {\vect V}_b(m', j'); \vect Y_b)> \gamma^{\U}]  \notag 
\end{IEEEeqnarray}
where  $m' \in \{1, \ldots, L_{\U}\}$, $j' \in \{1, \ldots, L_v\}$, $(M_b^{(\U)}, j) \neq (m',j')$,  $\bar{\vect V}_b \sim f_{\vect V_b}$ and is  independent of $(\vect V_b, \vect Y_b)$. 

\begin{lemma} \label{lemma3}
For any  $\gamma^{(\U)}>0$: 
\begin{IEEEeqnarray}{rCl}
\Pr[i(\bar {\vect V}_b; \vect Y_b)> \gamma^{(\U)}] \le e^{-\gamma^{(\U)}}.
\end{IEEEeqnarray}
\end{lemma}
\begin{IEEEproof}
See Appendix~\ref{App:C}.
\end{IEEEproof}

By Lemmas~\ref{lemma2} and~\ref{lemma3}, we have
\begin{IEEEeqnarray}{rCl} \label{eq:e2}
\lefteqn{\Pr [\mathcal E_{\U,3}| \mathcal E_{\U,2}^c, \mathcal E_{\U,1}^c, b \in \mb ]} \\
&\le&1- G\left(\frac{\Nu}{2}, \lambda (\mu_{\U}) \right) + G\left(\frac{\Nu}{2}, \tilde \lambda (\mu_{\U})\right)    + (L_vL_{\U} -1) e^{-\gamma^{(\U)}} .\notag 
\end{IEEEeqnarray}

\subsubsection{Analyzing $  \Pr [\mathcal E_{\U,4} | b \notin \mb]  $}
This error event is equivalent to the probability that given no URLLC is arrived, there exists at least one codeword $V_b(m,i)$ with $ m \in [L_{\U}]$ and $j \in [ L_v]$ such that $i(\vect V_b(m,j); \vect Y_b) > \gamma^{(\U)}$. Therefore, 
\begin{IEEEeqnarray}{rCl}
\lefteqn{  \Pr [\mathcal E_{\U,4} | b \notin \mb] } \notag \\
&=& 1 - \left (\Pr \left[i(\vect V_b(m,j); \vect Y_b) \le \gamma^{(\U)} \right]\right )^{L_vL_{\U}} \\
& \le & 1- \left (1- e^{-\gamma^{(\U)}} \right )^{L_vL_{\U}}. \label{eq:e4}
\end{IEEEeqnarray}
where the last inequality follows by Lemma~\ref{lemma3}. 

By combining \eqref{eq:e1}, \eqref{eq:e2}, \eqref{eq:e3} and \eqref{eq:e4} we prove bound \eqref{boundu}.

\subsection{Bounding $\epsilon^{(\e)}_{\text{TIN}}$}

Define 
\begin{subequations}\label{eq:rus}
\begin{IEEEeqnarray}{rCl} 
\ru &:=&  \Pr [b \in \mbt], \label{eq:ru} \\
\rdz&:=& \Pr[b \in \mbd| b \in \mbt], \label{eq:rdz}\\
\rdo&:=& \Pr[b \in \mbd| b \notin \mbt].  \label{eq:rdo}
\end{IEEEeqnarray}
\end{subequations}
\begin{lemma}\label{lemma4new}
We prove that 
\begin{IEEEeqnarray}{rCl}
\ru &=& \rho\left(1- (1-\zeta)^{L_v}\right), \quad \rdo \le q_1, \quad q_2 \le  \rdz \le q_3,\notag \\ \IEEEeqnarraynumspace
\end{IEEEeqnarray}
where $q_1$, $q_2$ and $q_3$ are defined in \eqref{eq:defs} and $\zeta$  in \eqref{eq:zeta}.
\end{lemma}
\begin{IEEEproof}
See Appendix~\ref{App:Dnew}. 
\end{IEEEproof}

 Given  $\mbd=\bku$, we have the following two error events:
\begin{IEEEeqnarray}{rCl}
\mathcal E_{\text{TIN},1} &=& \{\mbd \neq \mbt \} \\
\mathcal E_{\text{TIN},2} &=& \{\hat M^{(\e)} \neq M^{(\e)} \} .
\end{IEEEeqnarray}


The  eMBB decoding error probability under the TIN approach thus is bounded by
\begin{IEEEeqnarray}{rCl}
\lefteqn{\epsilon_{\e}^{\text{TIN}} \le \sum_{\bku} \Pr [\mbd = \bku]} \\
&& \cdot \left ( \Pr [\mathcal E_{\text{TIN},1}| \mbd = \bku] +\Pr[ \mathcal E_{\text{TIN},2}| \mbd = \bku, \mathcal E_{\text{TIN},1}^c]\right).\notag 
\end{IEEEeqnarray}

\subsubsection{Analyzing $ \Pr [\mbd = \bku]$}

Define
\begin{IEEEeqnarray}{rCl}\label{eq:rd}
\rd &:=& \Pr[b \in \mbd, b \in \mbt] + \Pr [ b \in \mbd, b \notin \mbt]  \IEEEeqnarraynumspace\\
& =&  \ru \rdz + (1- \ru) \rdo,\IEEEeqnarraynumspace
\end{IEEEeqnarray}
where $\ru$, $\rdz$ and $\rdo$ are defined in \eqref{eq:rus}. By Lemma~\ref{lemma4new}:
\begin{IEEEeqnarray}{rCl}
\ru \cdot q_2 \le \rd \le \ru \cdot q_3 + (1- \ru)\cdot q_1,
\end{IEEEeqnarray}
and thus by the independence of the blocks: 
\begin{IEEEeqnarray}{rCl}
\lefteqn{\Pr [\mbd = \bku ]}\\
 &=& \rd^{|\bku| } (1 - \rd)^{\eta- |\bku|}\\
&\le& (\ru \cdot q_3 + (1- \ru)\cdot q_1)^{|\bku| } (1-\ru \cdot q_2)^{\eta- |\bku|}
\end{IEEEeqnarray}
%

\subsubsection{Analyzing $ \Pr [\mathcal E_{\text{TIN},1}| \mbd = \bku ]$}

Notice that the values of $\ru, \rdz$ and $\rdo$   stay the same for all blocks in $[\eta]$. Thus 
\begin{IEEEeqnarray}{rCl}
\lefteqn{\Pr [\mbd \neq \mbt| \mbd = \bku]} \\
&=& 1- \Pr[\mbt = \bku| \mbd = \bku ]  \\
& = & 1 - \frac{\Pr[\mbt = \bku, \mbd = \bku ]}{\Pr[\mbd = \bku ]}  \\
& = & 1- \frac{\Pr[\mbt = \bku] \Pr[\mbd = \bku| \mbt = \bku]}{\rd^{|\bku|} (1 - \rd)^{\eta- |\bku|}}\\
& = & 1- \frac{\ru^{|\bku|}(1-\ru)^{\eta-|\bku|} \rdz^{|\bku|} (1-\rdo)^{\eta-|\bku|}}{\rd^{|\bku|} (1 - \rd)^{\eta- |\bku|}} \nonumber \\ \\
& \le & 1 -  \frac{\ru^{|\bku|}(1-\ru)^{\eta-|\bku|} q_2^{|\bku|}  (1-q_1)^{\eta-|\bku|}}{(\ru \cdot q_3 + (1- \ru)\cdot q_1)^{|\bku| } (1-\ru \cdot q_2)^{\eta- |\bku|}}\notag\\ \label{eq:66}
\end{IEEEeqnarray}
where $\ru, q_1,q_2$ and $q_3$ are defined in \eqref{eq:defs}. 
The inequality in \eqref{eq:66} follows by Lemma~\ref{lemma4new}. 

\subsubsection{Analyzing $\Pr[ \mathcal E_{\text{TIN},2}| \mbd = \bku, \mathcal E_{\text{TIN},1}^c]$}
To bound $\Pr[\hat M^{(\e)} \neq M^{(\e)} |\mbd = \bku,  \mathcal E_{\text{TIN},1}^c ]$, we use the threshold bound for maximum-metric decoding. For any given threshold $\gamma^{(\e)}$:
\begin{IEEEeqnarray}{rCl}
\lefteqn{\Pr[\hat M^{(\e)} \neq M^{(\e)} | \mbd = \bku,  \mathcal E_{\text{TIN},1}^c ]}  \notag \\
& &\hspace{-0.5cm}\le \Pr \left [i^{(\e)}_{\text{TIN}} \left ( \{\xkes \}_{b \notin  \bku },  \{\xkef\}_{b \in \bku }; \vect Y^{\Ne} | \bku  \right) < \gamma^{(\e)}\right ]\notag \\
&+&   \Pr \left [i^{(\e)}_{\text{TIN}} \left ( \{\bar {\vect X}_{b}^{(\e,1)} \}_{b \notin  \bku },  \{\bar {\vect X}_{b}^{(\e,2)}\}_{b \in \bku };  \vect Y^{\Ne}| \bku \right) \ge \gamma^{(\e)}\right ] \notag \\
&&\cdot (L_{\e}-1)  \label{eq:381}
\end{IEEEeqnarray}
where for each $b$, $\bar {\vect X}_{b}^{(\e,1)} \sim f_{\xkes}$ and  $\bar {\vect X}_{b}^{(\e,2)} \sim f_{\xkef}$ and are independent of $(\xkes, \xkef, \vect Y_b)$. We use the following two lemmas to bound the above two probability terms.

\begin{lemma}\label{lemma4}
For any $\gamma^{(\e)} >0$:
\begin{IEEEeqnarray}{rCl}
&&\Pr \left [i^{(\e)}_{\text{TIN}} \left ( \{\xkes \}_{b \notin  \bku },  \{\xkef\}_{b \in \bku };   Y^{\Ne} | \bku  \right) < \gamma^{(\e)}\right ] \notag \\
&& \le T - (L_v-1)e^{-\gamma^{(\e)}}
\end{IEEEeqnarray}
where $T$ is defined in \eqref{eq:T1}. 
\end{lemma}
\begin{IEEEproof}
See Appendix~\ref{App:D}.
\end{IEEEproof}
\begin{lemma} \label{lemma5}
For any $\gamma^{(\e)}>0$:
\begin{IEEEeqnarray}{rCl}
&&\Pr \left [i^{(\e)}_{\text{TIN}} \left ( \{\bar {\vect X}_{b}^{(\e,1)} \}_{b \notin  \bku },  \{\bar {\vect X}_{b}^{(\e,2)}\}_{b \in \bku };  \{\vect Y_b\}_{b = 1}^{\eta+1} | \bku \right) \ge \gamma^{(\e)}\right ] \notag \\
& &\le  e^{-\gamma^{(\e)}}. 
\end{IEEEeqnarray}
\end{lemma}
\begin{IEEEproof}
The proof is similar to the proof of  Lemma~\ref{lemma3} and omitted.
\end{IEEEproof}
Combining Lemmas~\ref{lemma4} and~\ref{lemma5} with \eqref{eq:381} and defining $k:=|\bku|$ proves the bound in \eqref{boundeTIN}. 
 
\subsection{Bounding $\epsilon^{(\e)}_{\text{SIC}}$}


Recall the definition of the sets $\mb$, $\mbt$, $\mbd$ and $\mbc$ from \eqref{eq:mb}, \eqref{eq:mbt}, \eqref{eq:mbd}, and \eqref{eq:mbc}, respectively. Let $\bku$ be a realization of the set $\mbd$, and $\bkut$ be a realization of the set $\mbc$. We have the following two error events:
\begin{IEEEeqnarray}{rCl}
\mathcal E_{\text{SIC},1} &=& \{\mbd \neq \mbt \} \\
\mathcal E_{\text{SIC},2} &=& \{\hat M^{(\e)} \neq M^{(\e)} \} 
\end{IEEEeqnarray}
The  eMBB decoding error probability under the SIC approach  thus is given by
\begin{IEEEeqnarray}{rCl}
\epsilon_{\e}^{\text{SIC}}  
&& \le \sum_{\bku} \Pr [\mbd = \bku] \notag \\
&&\Big ( \Pr [\mathcal E_{\text{SIC},1}| \mbd = \bku] \notag \\
&&+\sum_{\bkut} \Pr[\mbc = \bkut| \mathcal E_{\text{SIC},1}^c, \mbd = \bku] \notag \\
&& \hspace{0.5cm}\cdot \Pr[ \mathcal E_{\text{SIC},2}| \mbd = \bku, \mbc = \bkut, \mathcal E_{\text{SIC},1}^c]\Big). \IEEEeqnarraynumspace
\end{IEEEeqnarray}
\subsubsection{Analyzing $\Pr[\mbc = \bkut| \mathcal E_{\text{SIC},1}^c, \mbd = \bku]$} 
For any subset $B_{c} \subseteq B_d$ we have: 
\begin{IEEEeqnarray}{rCl}
\lefteqn{\Pr[\mbc = \bkut| \mbd = \mbt= \bku]}\\
&=&  \prod_{b \in \bkut} \Pr [\hat M_b^{(\U)} = M_b^{(\U)}| \mbd = \mbt = \bku] \notag \\
&& \cdot \prod_{b\in \bku \backslash \bkut} \left (1- \Pr [\hat M_b^{(\U)} = M_b^{(\U)}| \mbd = \mbt=\bku]\right ) \\
&\le& q^{|\bkut|} (1- q)^{|\bku| - |\bkut|} \label{eq:78}
\end{IEEEeqnarray}
where $q$ is defined in \eqref{eq:defs}. 
Inequality \eqref{eq:78} holds by \eqref{eq:e2} and by the independence of the blocks.
\subsubsection{Analyzing $\Pr[ \mathcal E_{\text{SIC},2}| \mbd = \bku, \mbc = \bkut, \mathcal E_{\text{SIC},1}^c]$}
To bound this probability, we use the threshold bound for maximum-metric decoding. For any given threshold $\tilde \gamma^{(\e)}$:
\begin{IEEEeqnarray}{rCl} \label{eq:38}
\lefteqn{\Pr[\hat M^{(\e)} \neq M^{(\e)} | \mbd = \bku, \mbc = \bkut, \mathcal E_{\text{SIC},1}^c ]} \\
& \le& \Pr \Big [i^{(\e)}_{\text{SIC}}  ( \{\xkes \}_{b \notin  \bku },  \{\xkef\}_{b \in \bku }; \notag \\
&& \hspace{1cm} \vect Y^{\Ne}| \bku, \bkut,   \{\vect V_b\}_{b \in \bkut} ) < \tilde \gamma^{(\e)}\Big]\notag \\
&&+ (L_{\e}-1)    \Pr \Big [i^{(\e)}_{\text{SIC}}  ( \{\bar {\vect X}_{b,1}^{(\e)} \}_{b \notin  \bku },  \{\bar {\vect X}_{b,2}^{(\e)}\}_{b \in \bku }; \notag \\
&& \hspace{2cm}\vect Y_b^{\Ne} | \bku, \bkut,  \{\vect V_b\}_{b \in \bkut} ) \ge \tilde \gamma^{(\e)}\Big] \IEEEeqnarraynumspace
\end{IEEEeqnarray}
where for each $b$, $\bar {\vect X}_{b}^{(\e,1)} \sim f_{\xkes}$ and  $\bar {\vect X}_{b}^{(\e,2)} \sim f_{\xkef}$ and are independent of $(\xkes, \xkef, \vect Y^{\Ne})$. We use the following two lemmas to bound the above two probability terms.
\begin{lemma}\label{lemma6}
Given $\tilde \gamma^{(\e)}$, we prove that 
\begin{IEEEeqnarray}{rCl}
&&\Pr \Big [i^{(\e)}_{\text{SIC}}  ( \{\xkes \}_{b \notin  \bku },  \{\xkef\}_{b \in \bku }; \{\vect Y_b\}_{b = 1}^{\eta+1} \notag \\
&& \hspace{4cm} | \bku,  \{\vect V_b\}_{b \in \bku} ) < \tilde \gamma^{(\e)}\Big]\notag \\
&&\le  \frac{\mu T }{\tilde \mu} -\nu
\end{IEEEeqnarray}
where $T$, $\nu, \mu$ and $\tilde \mu$ are defined in \eqref{eq:37}.
\end{lemma}
\begin{IEEEproof}
See Appendix~\ref{App:E}.
\end{IEEEproof}
\begin{lemma} \label{lemma7}
 We can prove that 
\begin{IEEEeqnarray}{rCl}
&&\Pr \Big [i^{(\e)}_{\text{SIC}}  ( \{\bar {\vect X}_{b,1}^{(\e)} \}_{b \notin  \bku },  \{\bar {\vect X}_{b,2}^{(\e)}\}_{b \in \bku }; \notag \\
&& \hspace{1cm}\{\vect Y_b\}_{b = 1}^{\eta+1} | \bku,  \{\vect V_b\}_{b \in \bku} ) \ge \tilde \gamma^{(\e)}\Big]\le  e^{-\tilde \gamma^{(\e)}}. 
\end{IEEEeqnarray}
\end{lemma}
\begin{IEEEproof}
The proof is based on the argument provided in the proof of  Lemma~\ref{lemma3}.
\end{IEEEproof}
Combining Lemmas~\ref{lemma6} and~\ref{lemma7} with  \eqref{eq:78} and defining $\tilde k = |\bkut|$  proves the bound in \eqref{boundeSIC}.

\section{Conclusions}
We considered a point-to-point scenario where a roadside unite (RSU) wishes to simultaneously send eMBB and URLLC messages to a vehicle.  During each eMBB transmission interval, random arrivals of URLLC messages are assumed. To improve the reliability of the URLLC transmissions, we proposed a coding scheme that mitigates the interference of eMBB transmission by means of dirty paper coding (DPC).  
 We derived rigorous upper bounds on the error probabilities of eMBB and URLLC transmissions achieved by our scheme. Our numerical analysis shows that the proposed scheme significantly improves over
the standard time-sharing. 
\section*{Acknowledgment}
The work of   H. V. Poor has been supported by the U.S. National Science Foundation (NSF) within the Israel-US Binational program
under grant CCF-1908308.
The work of S. Shamai (Shitz) has been supported
by the US-Israel Binational Science Foundation
(BSF) under grant BSF-2018710.

\appendices
\section{Proof of Lemma~\ref{lemma1}} \label{App:A}
By \eqref{eq:25} and since  $ \xkef$ and $\vect{V}_b$ are  drawn uniformly on the  $\Nu$-dimensional spheres of radii $\sqrt{\Nu \beta_{\e}\P}$ and $\sqrt{\Nu (\beta_{\U}+\alpha^2 \beta_{\e})\P}$, the error event $\mathcal{E}_{b,v}$ holds whenever the following condition is violated: 
\begin{IEEEeqnarray}{rCl} \label{eq:45}
\alpha \bef \Nu \Pb \le  \langle\vect V_b,  \xkef \rangle \le \alpha \bef \Nu \Pb+ \frac{\delta_{b}}{2 \alpha}. \IEEEeqnarraynumspace
\end{IEEEeqnarray}
The distribution of  $\langle\vect V_b,   \xkef \rangle$  depends on $\vect V_b$ only through its magnitude, because $\xkef$ is uniform over a sphere and applying an orthogonal transformation to $\vect V_b$ and $\xkef$ does neither change  the inner product of the two vectors nor the distribution of $\xkef$. 
  In the following we therefore assume that $\vect V_b = (||\vect V_b||, 0, \ldots, 0)$, in which case \eqref{eq:45} is equivalent to: 
\begin{IEEEeqnarray}{rCl}
\frac{\alpha \bef \Nu \Pb}{\sqrt{\vnorm \Nu \Pb}} \le X_{b,2,1}^{(\e)} \le \frac{\alpha \bef \Nu \Pb}{\sqrt{\vnorm \Nu \Pb}} + \frac{\delta_{b}}{2 \alpha\sqrt{\vnorm \Nu \Pb}}
\end{IEEEeqnarray}
where $X_{b,2,1}^{(\e)}$ is the first entry of the vector $\xkef$.

The distribution of a given symbol in a length-$\Nu$ random sequence distributed uniformly on the sphere is \cite{Stam1982}
\begin{IEEEeqnarray}{rCl}\label{eq:fx}
f_{X_{b,2,1}^{(\e)}}\left(x_{b,2,1}^{(\e)}\right)&=& \frac{1}{\sqrt{\pi \Nu \bef \Pb}}\frac{\Gamma(\frac{\Nu}{2})}{\Gamma (\frac{\Nu-1}{2})} \left (1 - \frac{(x_{b,2,1}^{(\e)})^2}{\Nu \bef \Pb} \right)^{\frac{\Nu-3}{2}} \notag\\
&& \times \mathbbm{1}\{(x_{b,2,1}^{(\e)})^2 \le  \Nu \bef \Pb \}.
\end{IEEEeqnarray}
Thus,
\begin{IEEEeqnarray}{rCl} \label{eq:pu}
\lefteqn{\Pr \left[\vect V_b - \alpha  \xkef  \in \mathcal D_k \right] } \notag \\
&=& \int_{\frac{\alpha \bef \Nu \Pb}{\sqrt{\vnorm \Nu \Pb}}}^{\frac{\alpha \bef \Nu \Pb}{\sqrt{\vnorm \Nu \Pb}} + \frac{\delta_{b}}{2 \alpha\sqrt{\vnorm \Nu \Pb}}}   f_{X_{b,2,1}^{(\e)}}\left(x_{b,2,1}^{(\e)} \right) dx_{b,2,1}^{(\e)} \\
& = & \frac{1}{\sqrt{\pi }}\frac{\Gamma(\frac{\Nu}{2})}{\Gamma (\frac{\Nu-1}{2})} \kappa_{\frac{\Nu-3}{2}} \left ( \frac{2 \alpha^2 \Nu \Pb \bef + \delta_{b}}{2 \alpha \Nu \Pb \sqrt{\vnorm\bef}} \right ) \notag \\
&& - \frac{1}{\sqrt{\pi }}\frac{\Gamma(\frac{\Nu}{2})}{\Gamma (\frac{\Nu-1}{2})} \kappa_{\frac{\Nu-3}{2}} \left (  \alpha\sqrt{\frac{\bef}{\vnorm}}\right ) ,
\end{IEEEeqnarray}
where
\begin{IEEEeqnarray}{rCl}
\kappa_n(x) = \frac{x(1-x^2)^n}{2n+1} + \frac{2n}{2n+1} \kappa_{n-1}(x)
\end{IEEEeqnarray}
with $\kappa_0(x) = x$. 
This concludes the proof. 

\section{Proof of Lemma~\ref{lemma2}} \label{App:B}
Note that $\vect Y_b$  and $\vect Y_b| \vect V_b$ do not follow a Gaussian distribution. 
Define 
\begin{IEEEeqnarray}{rCl}
Q_{\vect Y_b} (\vect y_b) &=& \mathcal N(\vect y_{k,1}; \vect 0, I_{\Nu} \sy) \label{eq:qy}\\
  Q_{\vect Y_b| \vect V_b} (\vect y_b| \vect v_b) &=& \mathcal N(\vect y_h; h \vect V_b, I_{\Nu} \syv) \label{eq:qyv}
\end{IEEEeqnarray}
with $\sy = h^2  \Pb + 1$ and $\syv = h^ 2(1 - \alpha)^2 \bef \Pb+ 1$. 

Introduce 
\begin{IEEEeqnarray}{rCl}
\tilde i_b^{(\U)}(\vect v_b; \vect y_b ) := \ln \frac{Q_{\vect Y_b| \vect V_b} (\vect y_b| \vect v_b) }{Q_{\vect Y_b} (\vect y_b)  }.
\end{IEEEeqnarray}
\begin{lemma} \label{lemmaJU}
We can prove that 
\begin{IEEEeqnarray}{rCl}
 i_b^{(\U)}(\vect v_b; \vect y_b ) \ge \tilde i_b^{(\U)}(\vect v_b; \vect y_b ) + \ln  J_{\U},
\end{IEEEeqnarray}
where
\begin{IEEEeqnarray}{rCl} \label{eq:ju}
J_{\U} & :=& \frac{ \pi \sqrt{ \vnorm \bef}  2^{\frac{\Nu+1}{2}}e^{-\frac{h^2(1-\alpha)^2\bef \Pb \Nu}{2}}}{9h^2(1-\alpha) (\vnorm + (1-\alpha)^2 \bef  )}
\end{IEEEeqnarray}
\end{lemma}
\begin{IEEEproof}
By \cite[Propsition 2]{Molavianjazi}: 
\begin{IEEEeqnarray}{rCl}
\frac{f_{\vect Y_b} (\vect y_b)}{Q_{\vect Y_b} (\vect y_b)} \le  \frac{9((1-\alpha)h)^{\Nu}}{2 \pi \sqrt{2}} \frac{\vnorm \Pb + (1-\alpha)^2 \bef \Pb }{(1-\alpha) \Pb\sqrt{ \vnorm \bef}}.
\end{IEEEeqnarray}
By \cite[Lemma 5]{Nikbakht2022}:
\begin{IEEEeqnarray}{rCl}
\frac{f_{\vect Y_b| \vect V_b} (\vect y_b| \vect v_b)}{Q_{\vect Y_b| \vect V_b} (\vect y_b| \vect v_b)} \ge 2^{\frac{\Nu-2}{2}}\left(h(1-\alpha)\right)^{\Nu -2} e^{-\frac{h^2(1-\alpha)^2\bef \Pb \Nu}{2}}\IEEEeqnarraynumspace
\end{IEEEeqnarray}
Combining the two bounds concludes the proof. 
\end{IEEEproof}
As a result, we have 
\begin{IEEEeqnarray}{rCl}
  \lefteqn{  \Pr [i_b^{(\U)}(\vect V_b; \vect Y_b) \le \gamma^{(\U)} ]} \\
& \le & \Pr [\tilde i(\vect V_b; \vect Y_b ) \le \gamma^{(\U)}  - \ln J_{\U}] \\
& = &    \Pr  \left [\ln \frac{Q_{\vect Y_b| \vect V_b}(\vect Y_b| \vect V_b)}{Q_{\vect Y_b}(\vect Y_b) }  \le \gamma^{(\U)}  - \ln J_{\U} \right ] \\
& = & \Pr \Bigg [ \ln {\frac{\frac{1}{(\sqrt{2\syv\pi})^{\Nu}}\exp \left (- \frac{|| \vect Y_b - h\vect V_b||^2}{2\syv}\right )}{\frac{1}{(\sqrt{2\pi \sy})^{\Nu}}\exp \left (- \frac{|| \vect Y_b||^2}{2 \sy}\right )}} \le \gamma^{(\U)}  - \ln J_{\U}\Bigg ] \\
&= & \Pr \Bigg [ \frac{\Nu}{2} \ln \frac{\sy}{\syv} + \frac{|| \vect Y_b||^2}{2 \sy}- \frac{|| \vect Y_b - h\vect V_b||^2}{2\syv} \le \gamma^{(\U)}  - \ln J_{\U}\Bigg ] \\
& = & \Pr \Bigg [\frac{h^2}{2\sy}|| \xku ||^2 + \frac{h^2}{2}\left (\frac{1}{\sy} - \frac{(1-\alpha)^2}{\syv}\right ) ||\xkef||^2    \notag \\
&& \hspace{0.3cm} +\frac{h^2}{2}\left (\frac{1}{\sy} - \frac{1}{\syv}\right ) ||\vect Z_b||^2+ \frac{h}{\sy} \langle \xku, \xkef \rangle \notag \\
&& \hspace{0.5cm} + \frac{h}{\sy} \langle \xku, \vect Z_b \rangle  + \left (\frac{h}{\sy} + \frac{h(1-\alpha)}{\syv} \right) \langle \xkef, \vect Z_b \rangle  \notag \\
&&\hspace{4cm} \le \gamma^{(\U)}  - \ln J_{\U} - \frac{\Nu}{2}\ln \frac{\sy}{\syv} \Bigg ] \IEEEeqnarraynumspace \\
& \le & \Pr \Bigg [\frac{h^2(\Nu \bu \Pb - \delta_{b})}{2\sy} + \frac{h^2\Nu \bef \Pb}{2}\left (\frac{1}{\sy} - \frac{(1-\alpha)^2}{\syv}\right )    \notag \\
&& \hspace{0.3cm} +\frac{h^2}{2}\left (\frac{1}{\sy} - \frac{1}{\syv}\right ) ||\vect Z_b||^2 -\frac{h \Nu \Pb \sqrt{\bu \bef}}{\sy} \notag \\
&& \hspace{0.5cm} - h \sqrt{\Nu \Pb} \left (\frac{\sqrt{\bu}}{\sy}  + \frac{\sqrt{\bef}}{\sy} + \frac{\sqrt{\bef} (1-\alpha)}{\syv} \right ) || \vect Z_b|| \notag \\
&&\hspace{4.2cm} \le \gamma^{(\U)}  - \ln J_{\U} - \frac{\Nu}{2}\ln \frac{\sy}{\syv} \Bigg ] \IEEEeqnarraynumspace \\
& = & \Pr \left  [ ||\vect Z_b||^2 + u ||\vect Z_b|| \ge \mu_{\U}\right ] \\
& = & \Pr \left  [ \left (||\vect Z_b|| + \frac{u}{2} \right ) ^2  \ge \mu_{\U} + \frac{u^2}{4}\right ] \\
& = & 1- F\left (\sqrt{\mu_{\U} + \frac{u^2}{4}} - \frac{u}{2}\right ) + F \left (-\sqrt{\mu_{\U} + \frac{u^2}{4}} - \frac{u}{2} \right ) \label{eq:103}\IEEEeqnarraynumspace
\end{IEEEeqnarray}
where 
\begin{subequations}
\begin{IEEEeqnarray}{rCl}
 \mu_{\U} & := & \frac{2\sy \syv}{h^2(\sy-\syv)} \left(\frac{\Nu}{2}\ln \frac{\sy}{\syv} - \gamma^{(\U)}  + \ln J_{\U} \right)  \notag \\
&& + \frac{\syv}{\sy - \syv} \left ( \Nu \Pb (\sqrt{\bu} - \sqrt{\bef})^2 - \delta_b\right) \notag  \\
&& -  \frac{\sy \Nu \bef \Pb (1-\alpha)^2}{\sy - \syv} \notag\\
u & := &   \frac{2\sqrt{\Nu \Pb} \left ( \syv (\sqrt{\bu} + \sqrt{\bef}) + \sy \sqrt{\bef}(1- \alpha)\right)}{h (\sy - \syv)} \notag
\end{IEEEeqnarray}
\end{subequations}
Notice that in \eqref{eq:103} we use the fact that $||\vect Z_b||$ follows a chi-distribution with degree $\Nu$ and $F(\cdot)$ represents its CDF.  
\section{Proof of Lemma~\ref{lemma3}}\label{App:C}
By Bayes' rule we have
\begin{IEEEeqnarray}{rCl}\label{eq:44}
f_{\vect V_b}(\bar {\vect v}_b) &=& \frac{f_{\vect Y_b}(\vect y_b)f_{\vect V_b | \vect Y_b} (\bar {\vect v}_b | \vect y_b) }{f_{\vect Y_b| \vect V_b} (\vect y_b| \bar {\vect v}_b)}\\
& =& f_{\vect V_b | \vect Y_b} (\bar {\vect v}_b | \vect y_b) \exp \left ( - i(\bar{\vect v}_b, \vect y_b ) \right ). 
\end{IEEEeqnarray}
By multiplying both sides of the above equation by $\mathbbm {1} \{i(\bar{\vect v}_b, \vect y_b )> \gamma\}$ and integrating over all $\bar{\vect v}_b$, we have
\begin{IEEEeqnarray}{rCl} \label{eq:63}
\lefteqn{\int_{\bar {\vect v}_b} \mathbbm {1} \{i(\bar{\vect v}_b, \vect y_b )> \gamma\} f_{\vect V_b}(\bar {\vect v}_b) d \bar {\vect v}_b = } \notag \\
& &  \int_{\bar {\vect v}_b} \mathbbm {1} \{i(\bar{\vect v}_b, \vect y_b )> \gamma\}  e^{  - i(\bar{\vect v}_b, \vect y_b )} f_{\vect V_b| \vect Y_b} (\bar {\vect v}_b | \vect y_b) d \bar {\vect v}_b. \IEEEeqnarraynumspace
\end{IEEEeqnarray}
Note that the left-hand side of \eqref{eq:63} is equivalent to $\Pr [i(\bar{\vect v}_b, \vect y_b)> \gamma| \vect Y_b = \vect y_b ] $. Thus 
\begin{IEEEeqnarray}{rCl}
\lefteqn{\Pr [i(\bar{\vect v}_b, \vect y_b )> \gamma| \vect Y_b = \vect y_b ] } \\
&= & \int_{\bar {\vect v}_b} \mathbbm {1} \{i(\bar{\vect v}_b, \vect y_b )> \gamma\}  \notag \\
&& \hspace{0.5cm} \times \exp \left ( - i(\bar{\vect v}_b, \vect y_b ) \right ) f_{\vect V_b | \vect Y_b} (\bar {\vect v}_b | \vect y_b) d \bar {\vect v}_b \\
& = & \int_{\bar {\vect v}_b} \mathbbm {1} \left \{\frac{f_{\vect Y_b| \vect V_b} (\vect y_b| \bar {\vect v}_b)}{f_{\vect Y_b}(\vect y_b)} e^{-\gamma}>1 \right\}  \notag \\
&& \hspace{0.5cm} \times \exp \left ( - i(\bar{\vect v}_b, \vect y_b ) \right ) f_{\vect V_b | \vect Y_b} (\bar {\vect v}_b | \vect y_b) d \bar {\vect v}_b \\
& \le & \int_{\bar {\vect v}_b} \frac{f_{\vect Y_b| \vect V_b}  (\vect y_b| \bar {\vect v}_b)}{f_{\vect Y_b}(\vect y_b)} e^{-\gamma}  \notag \\
&& \hspace{0.5cm} \times \exp \left ( - i(\bar{\vect v}_b, \vect y_b ) \right ) f_{\vect V_b | \vect Y_b} (\bar {\vect v}_b | \vect y_b) d \bar {\vect v}_b \\
& = & \int_{\bar {\vect v}_b} e^{-\gamma} f_{\vect V_b | \vect Y_b} (\bar {\vect v}_b | \vect y_b) d \bar {\vect v}_b \\
& = &  e^{-\gamma}. 
\end{IEEEeqnarray}
\section{Proof of Lemma~\ref{lemma4new}}\label{App:Dnew}
We start by analyzing the quantities in $\ru$, $\rdz$ and $\rdo$ defined in \eqref{eq:ru}, \eqref{eq:rdz} and \eqref{eq:rdo}.
\subsubsection{Analyzing $\ru$}
\begin{IEEEeqnarray}{rCl} \label{eq:ru-new}
\ru &=& \rho \cdot \Pr[ \exists \;  j \in [L_v] \; \text{s.t.}  \; \xku(\vect V_b(M_b^{(\U)},j) ) \in \mathcal D_b ] \IEEEeqnarraynumspace \\
& = & \rho (1- (1-\zeta)^{L_v})
\end{IEEEeqnarray}
where the last equality is by \eqref{eq:e1}. 
\subsubsection{Bounding $\rdz$} 
\begin{IEEEeqnarray}{rCl}
\lefteqn{\rdz }\notag \\
&=&  \Pr[b \in \mbd | b \in \mbt] \\
&= & 1- \Pr [\forall m, \forall j: i^{(\U)}_b  (\vect V_b(m,j); \vect Y_b ) \le \gamma^{(\U)}| b \in \mbt] \IEEEeqnarraynumspace\\
&\ge & 1 - \left  (1- G\left(\frac{\Nu}{2}, \lambda (\mu_{\U}) \right) + G\left(\frac{\Nu}{2}, \tilde \lambda (\mu_{\U})\right)  \right) ^{L_vL_{\U}} \label{eq:140}
\end{IEEEeqnarray}
where \eqref{eq:140} is by \eqref{eq:e3}. 

\begin{lemma} \label{lemma10}
For any $\gamma^{(\U)}>0$: 
\begin{IEEEeqnarray}{rCl}
\lefteqn{\Pr [i_b^{(\U)}(\vect V_b(m,j); \vect Y_b) \le \gamma^{(\U)} ] } \nonumber \\
&\ge& 1- G\left(\frac{\Nu}{2}, \tilde \lambda (\tilde \mu_{\U})\right) + G\left(\frac{\Nu}{2},  \lambda (\tilde \mu_{\U})\right)
\end{IEEEeqnarray}
where $G(\cdot,\cdot)$ is the regularized gamma function, $\lambda (\cdot)$ and $\tilde \lambda (\cdot)$ are defined in \eqref{eq:31} and $\tilde \mu_{\U}$ is defined in \eqref{eq:37}. 
\end{lemma}
\begin{IEEEproof}
The proof is similar to the proof of Lemma~\ref{lemma2}. We present a sketch of the proof. 

We start by upper bounding 
\begin{IEEEeqnarray}{rCl}
 i_b^{(\U)}(\vect v_b; \vect y_b )  \le \tilde i_b^{(\U)}(\vect v_b; \vect y_b ) +  \ln \tilde J_{\U},
\end{IEEEeqnarray}
where by \cite[Propsition 2]{Molavianjazi} and \cite[Lemma 6]{Nikbakht2022} we can prove that
\begin{IEEEeqnarray}{rCl} \label{eq:jut}
\tilde J_{\U} & :=&\frac{27 \sqrt{\pi} (1+h^2(1-\alpha)^2\bef \P)e^{\Nu h^2 \P (\vnorm + (1-\alpha)^2 \bef)}}{2(h^2(1-\alpha))^{\Nu-2}\sqrt{8 (1+2h^2(1-\alpha)^2\bef\Pb} }.\IEEEeqnarraynumspace
\end{IEEEeqnarray}

Thus 
\begin{IEEEeqnarray}{rCl}
  \lefteqn{  \Pr [i_b^{(\U)}(\vect V_b; \vect Y_b) \le \gamma^{(\U)} ]} \\
& \ge & \Pr [\tilde i(\vect V_b; \vect Y_b ) \le \gamma^{(\U)} - \ln \tilde J_{\U}] \\
& = & \Pr \left  [ ||\vect Z_b||^2 - u ||\vect Z_b|| \ge \tilde \mu_{\U}\right ] \\
& = & \Pr \left  [ \left (||\vect Z_b|| - \frac{u}{2} \right ) ^2  \ge \tilde \mu_{\U} + \frac{u^2}{4}\right ] \\
& = & 1- F\left (\sqrt{\tilde \mu_{\U} + \frac{u^2}{4}} + \frac{u}{2}\right ) + F \left (-\sqrt{\mu_{\U} + \frac{u^2}{4}} + \frac{u}{2} \right ) \IEEEeqnarraynumspace
\end{IEEEeqnarray}
\end{IEEEproof}
where 
\begin{IEEEeqnarray}{rCl}
\tilde  \mu_{\U} & := & \frac{2\sy \syv}{h^2(\sy-\syv)} \left(\frac{\Nu}{2}\ln \frac{\sy}{\syv} - \gamma^{(\U)} + \ln \tilde J_{\U} \right)  \notag \\
&& + \frac{\syv}{\sy - \syv} \left ( \Nu \Pb (\sqrt{\bu} +\sqrt{\bef})^2 \right) \notag  \\
&& -  \frac{\sy \Nu \bef \Pb (1-\alpha)^2}{\sy - \syv} 
\end{IEEEeqnarray}
By Lemma~\ref{lemma10}: 
\begin{IEEEeqnarray}{rCl}
\rdz \le 1 - \left  (1- G\left(\frac{\Nu}{2}, \tilde \lambda_1\right) + G \left(\frac{\Nu}{2}, \tilde \lambda_2 \right) \right) ^{L_vL_{\U}}.
\end{IEEEeqnarray}
\subsubsection{Upper  Bounding $\rdo$}
\begin{IEEEeqnarray}{rCl}
\lefteqn{\rdo }\notag \\
&=&  \Pr[b \in \mbd | b \notin \mbt] \\
& = & \Pr[\exists m \in [L_{\U}], j\in [L_v]: i^{(\U)}_b  (\vect V_b(m,j); \vect Y_b ) \ge \gamma^{(\U)}| b \notin \mbt] \notag\\
&= & 1- \Pr [\forall m, \forall j: i^{(\U)}_b  (\vect V_b(m,j); \vect Y_b ) \le \gamma^{(\U)}| b \in \mbt]\\
&\le & 1 - \left  (1- e^{-\gamma^{(\U)}} \right) ^{L_vL_{\U}} \label{eq:143}
\end{IEEEeqnarray}
where \eqref{eq:143} is by \eqref{eq:e4}. 
\section{Proof of Lemma~\ref{lemma4}}\label{App:D}
Notice that for each $b \in [1:\eta+1]$, $\vect Y_b$  and for $b \in \bku$, $\vect Y_b| \xkef$  do not follow a Gaussian distribution. 
Define $Q_{\vect Y_b}(\vect y_b)$ as in \eqref{eq:qy} and 
\begin{IEEEeqnarray}{rCl}
  Q_{\vect Y_b| \xkef} (\vect y_b| \vect x_{b}^{(\e,2)}) &=& \mathcal N(\vect y_b; h (1-\alpha)\xkef, I_{\Nu} \syx) \label{eq:qyx} \IEEEeqnarraynumspace
\end{IEEEeqnarray}
with  $\syx = h^ 2\vnorm \P+ 1$. 

Introduce 
\begin{IEEEeqnarray}{rCl}
\lefteqn{\tilde i^{(\e)}_{\text{TIN}} \left ( \{\vect x_{b}^{(\e,1)}\}_{b \notin  \bku },  \{\vect x_{b}^{(\e,2)}\}_{b \in \bku };  \{\vect y_b\}_{b = 1}^{\eta+1}|\bku \right)}\notag \\ &&
: = \ln \hspace{-0.15cm}\prod_{b\notin  \bku  }\hspace{-0.15cm} \frac{f_{\vect Y_b| \xkes} (\vect y_b| \vect x_{b}^{(\e,1)})}{Q_{\vect Y_b}(\vect y_b)} +  \ln \hspace{-0.15cm}\prod_{b\in  \bku } \hspace{-0.15cm} \frac{Q_{\vect Y_b| \xkef} (\vect y_b| \vect x_{b}^{(\e,2)})}{Q_{\vect Y_b}(\vect y_b)} \IEEEeqnarraynumspace
\end{IEEEeqnarray}
\begin{lemma} \label{lemmaJe}
We can prove that 
\begin{IEEEeqnarray}{rCl}
 \lefteqn{i^{(\e)}_{\text{TIN}} \left ( \{\vect x_{b}^{(\e,1)}\}_{b \notin  \bku },  \{\vect x_{b}^{(\e,2)}\}_{b \in \bku };  \{\vect y_b\}_{b = 1}^{\eta+1} | \bku\right)} \notag \\
&\ge& {\tilde i^{(\e)}_{\text{TIN}} \left ( \{\vect x_{b}^{(\e,1)}\}_{b \notin  \bku },  \{\vect x_{b}^{(\e,2)}\}_{b \in \bku };  \{\vect y_b\}_{b = 1}^{\eta+1} | \bku \right) }  + \ln J_{\e}, \IEEEeqnarraynumspace
\end{IEEEeqnarray}
where
\begin{IEEEeqnarray}{rCl} \label{eq:ju}
J_{\e} & :=& \left( \frac{\pi 2^{\frac{\Nu+1}{2}}e^{\frac{-h^2 \vnorm \Pb \Nu}{2}} \sqrt{\vnorm \bef}}{9h^2 (1- \alpha)^{\Nu -1} (\vnorm + (1- \alpha)^2 \bef )} \right)^k \notag \\
&& \cdot  \left ( \frac{\sqrt{8 (1 + 2 h^2\Pb)}}{27\sqrt{\pi} (1+ h^2 \Pb)} \right)^{\eta-k}
\end{IEEEeqnarray}
\end{lemma}
\begin{IEEEproof}
similar to the proof of Lemma~\ref{lemmaJU} and by \cite[Proposition 2]{Molavianjazi}, for $b \notin \bku$:
\begin{IEEEeqnarray}{rCl}
\frac{f_{\vect Y_b} (\vect y_b)}{Q_{\vect Y_b} (\vect y_b)} \le  \frac{27\sqrt{\pi} (1+h^2 \Pb)}{\sqrt{8 (1 + 2h^2 \Pb)}}.
\end{IEEEeqnarray}
\end{IEEEproof}
As a result, we have 
\begin{IEEEeqnarray}{rCl}
\lefteqn{\Pr \left [i^{(\e)}_{\text{TIN}} \left ( \{\xkes\}_{b \notin  \bku },  \{\xkef\}_{b \in \bku };  \vect Y^{\Ne} |\bku  \right)  < \gamma^{(\e)} \right] }\notag \\
&\le& \Pr \Big [\tilde i^{(\e)}_{\text{TIN}} \left ( \{\xkes\}_{b \notin  \bku },  \{\xkef\}_{b \in \bku };  \vect Y^{\Ne} |\bku  \right) \notag \\
&& \hspace{4.9cm} <\gamma^{(\e)} - \ln J_{\e} \Big] \notag\\
&=& \Pr \Bigg[ \ln \hspace{-0.15cm}\prod_{b\notin  \bku  }\hspace{-0.15cm} \frac{f_{\vect Y_b| \xkes} (\vect y_b| \vect x_{b}^{(\e,1)})}{Q_{\vect Y_b}(\vect y_b)} \notag \\
&& \hspace{0.75cm} +  \ln \hspace{-0.15cm}\prod_{b\in  \bku } \hspace{-0.15cm} \frac{Q_{\vect Y_b| \xkef} (\vect y_b| \vect x_{b}^{(\e,2)})}{Q_{\vect Y_b}(\vect y_b)}<\gamma^{(\e)} - \ln J_{\e} \Bigg] \notag \\
&=&\Pr \Bigg[ \ln \hspace{-0.25cm}\prod_{b\notin  \bku\backslash \eta+1  }\hspace{-0.25cm} \frac{\frac{1}{(\sqrt{2\pi})^{\Nu}} e^{-\frac{||\vect Z_b||^2}{2}}}{\frac{1}{(\sqrt{2\pi \sy})^{\Nu}} e^{-\frac{||\xkes + \vect Z_b||^2}{2 \sy}}} \notag \\
&& \hspace{0.25cm} +  \ln \hspace{-0.15cm}\prod_{b\in  \bku } \hspace{-0.15cm} \frac{\frac{1}{(\sqrt{2\pi \syx})^{\Nu}} e^{-\frac{||\vect V_b + \vect Z_b||^2}{2\syx}}}{\frac{1}{(\sqrt{2\pi \sy})^{\Nu}} e^{-\frac{||\xku + \xkef + \vect Z_b||^2}{2 \sy}}}\notag \\
&&  \hspace{0.25cm}+ \ln \frac{\frac{1}{(\sqrt{2\pi})^{\Ne-\eta\Nu}} e^{-\frac{|| \vect Z_{\eta+1}||^2}{2}}}{\frac{1}{(\sqrt{2\pi \sy})^{\Ne-\eta\Nu}} e^{-\frac{|| \vect X_{\eta+1,1}^{(\e)} + \vect Z_{\eta+1}||^2}{2 \sy}}}<\gamma^{(\e)} - \ln J_{\e} \Bigg]  \notag\\
&=&\Pr \Bigg[  \frac{1}{2} \sum_{b \notin  \bku} ||\vect Z_b||^2 - \frac{1}{2\sy} ||\xkes + \vect Z_b||^2 \notag \\
&& +  \sum_{b \in  \bku}\frac{||\vect V_b + \vect Z_b||^2}{2\syx}  -  \frac{ ||\vect V_b + (1- \alpha)\xkef + \vect Z_b||^2}{2\sy} \notag \\
&&>-\gamma^{(\e)} +\ln J_{\e} +\frac{\Ne}{2} \ln \sy  -\frac{\Nu k}{2}  \ln \syx \Bigg] \\
&\le&  \Pr \Bigg[ \frac{\sy-1}{2\sy} \sum_{b \notin  \bku} ||\vect Z_b||^2 +   \frac{\sqrt{\Nu \Pb}}{\sy}\sum_{b \notin  \bku} ||\vect Z_b|| \notag \\
&&+   \tau \sum_{b \in  \bku} ||\vect Z_b||  +  \frac{\sy-\syx}{2\sy \syx}\sum_{b \in  \bku} ||\vect Z_b||^2  > \mu \Bigg]\\
&\stackrel{{(a)}}{=}& \Pr \Bigg[ \frac{\sy-1}{2\sy} \tilde Z_1 +   \frac{\sqrt{\Nu \Pb}}{\sy}\sum_{b \notin  \bku} ||\vect Z_b|| \notag \\
&&+   \tau \sum_{b \in  \bku} ||\vect Z_b||  +  \frac{\sy-\syx}{2\sy \syx} \tilde Z_2  > \mu \Bigg]\\
&\stackrel{{(b)}}{\le}& \frac{\mathbb E \left [ \frac{\sy-1}{2\sy} \tilde Z_1 +   \frac{\sqrt{\Nu \Pb}}{\sy}\sum_{b \notin  \bku} ||\vect Z_b|| \right ]} {\mu}\notag \\
&&+ \frac{\mathbb E \left [  \tau \sum_{b \in  \bku} ||\vect Z_b||  +  \frac{\sy-\syx}{2\sy \syx} \tilde Z_2 \right]}{\mu} \\
&=& \frac{(\Ne - k \Nu)(\sy-1)}{2\sy \mu} +  \frac{(\eta+1 - k)\sqrt{\Nu \Pb}}{\sy \mu} \frac{\sqrt{2} \Gamma\left (\frac{\Nu +1}{2}\right)}{\Gamma\left (\frac{\Nu}{2}\right)} \notag \\
&& + \frac{k\tau}{\mu}\frac{\sqrt{2} \Gamma\left (\frac{\Nu +1}{2}\right)}{\Gamma\left (\frac{\Nu}{2}\right)} +  \frac{k \Nu (\sy-\syx)}{2\sy \syx \mu}
\end{IEEEeqnarray}
where
\begin{IEEEeqnarray}{rCl}
\tau&: =& \frac{\sqrt{\Nu \P} \left(\sqrt{\vnorm} (\sy + \syx) + (1- \alpha) \sqrt{\bef}\syx \right)}{\sy \syx}\notag \\
\mu&:=& -\gamma^{(\e)} + \ln J_{\e}+ \frac{\Ne}{2} \ln \sy - \frac{k\Nu}{2} \ln \syx - \frac{\eta+1-k}{2\sy} \Nu \Pb \notag \\
&& +   \frac{k}{2\syx}\vnorm \Nu \Pb - \frac{k}{2\sy}\left (\sqrt{\vnorm} + (1- \alpha)\sqrt{\bef} \right)^2\Nu \Pb \notag
\end{IEEEeqnarray}
In step $(a)$, we define 
\begin{IEEEeqnarray}{rCl}
\tilde Z_1 &:=& \sum_{b \notin  \bku} ||\vect Z_b||^2 \sim \mathcal X^2(\Ne - k\Nu) \\ 
\tilde Z_2 &:=& \sum_{b \in  \bku} ||\vect Z_b||^2 \sim \mathcal X^2(k\Nu)
\end{IEEEeqnarray}
where $\mathcal X^2(n)$ represents chi-squared distribution of degree $n$. 
In step $(b)$, we use the following  Markov's inequality:
\begin{IEEEeqnarray}{rCl}
\Pr [X > a] \le \frac{\mathbb E[X]}{a}.
\end{IEEEeqnarray}
In step $(c)$: 
\begin{IEEEeqnarray}{rCl}
\mathbb E [\tilde Z_1] &=& \Ne - k\Nu, \\
\mathbb E [\tilde Z_2] &=& k\Nu, \\
\mathbb E [||\vect Z_b||] &=& \frac{\sqrt{2} \Gamma\left (\frac{\Nu +1}{2}\right)}{\Gamma\left (\frac{\Nu}{2}\right)}.
\end{IEEEeqnarray}

\section{Proof of Lemma~\ref{lemma6}}\label{App:E}
Define $Q_{\vect Y_b} (\vect y_b)$ as in \eqref{eq:qy}, $Q_{\vect Y_b |\vect V_b} (\vect y_b| \vect v_b)$ as in \eqref{eq:qyv} and $Q_{\vect Y_b| \xkef} (\vect y_b| \vect x_{b}^{(\e,2)})$ as in \eqref{eq:qyx}.

Introduce 
\begin{IEEEeqnarray}{rCl}
\lefteqn{\tilde i^{(\e)}_{\text{SIC}} \Big( \{\vect x_{b}^{(\e,1)}\}_{b \notin \bku  },  \{\vect x_{b}^{(\e,2)}\}_{b \in \bku }; \vect y^{\Ne} | \bku , \bkut, \{\vect V_b\}_{b \in \bkut}\Big)} \notag \\ &&
: = \ln \hspace{-0.15cm}\prod_{b\notin \bku  }\hspace{-0.15cm} \frac{f_{\vect Y_b| \xkes} (\vect y_b| \vect x_{b}^{(\e,1)})}{Q_{\vect Y_b}(\vect y_b)} +  \ln \hspace{-0.15cm}\prod_{b\in  \bku \backslash \bkut } \hspace{-0.15cm} \frac{Q_{\vect Y_b| \xkef} (\vect y_b| \vect x_{b}^{(\e,2)})}{Q_{\vect Y_b}(\vect y_b)} \notag \\
&&+  \ln \hspace{-0.15cm}\prod_{b\in  \bkut } \hspace{-0.15cm} \frac{f_{\vect Y_b| \xkef, \vect V_b} (\vect y_b| \vect x_{b}^{(\e,2)}, \vect v_b)}{Q_{\vect Y_b| \vect V_b}(\vect y_b| \vect v_b)}
\end{IEEEeqnarray}
\begin{lemma} \label{lemmaJe2}
We can prove that 
\begin{IEEEeqnarray}{rCl}
 \lefteqn{i^{(\e)}_{\text{SIC}} \Big( \{\vect x_{b}^{(\e,1)}\}_{b \notin \bku  },  \{\vect x_{b}^{(\e,2)}\}_{b \in \bku }; \vect y^{\Ne} | \bku, \bkut, \{\vect v_b\}_{b \in \bkut}\Big)} \notag \\
&\ge&{\tilde i^{(\e)}_{\text{SIC}} \Big( \{\vect x_{b}^{(\e,1)}\}_{b \notin \bku  },  \{\vect x_{b}^{(\e,2)}\}_{b \in \bku }; \vect y^{\Ne} | \bku, \bkut,  \{\vect v_b\}_{b \in \bkut}\Big)} \notag \\
&&+ \ln \tilde J_{\e}, 
\end{IEEEeqnarray}
where
\begin{IEEEeqnarray}{rCl} \label{eq:ju}
\tilde J_{\e} & :=& \left ( \frac{\pi 2^{\frac{\Nu+1}{2}}e^{\frac{-h^2 \vnorm \Pb \Nu}{2}} \sqrt{\vnorm \bef}}{9h^2 (1- \alpha)^{\Nu -1} (\vnorm + (1- \alpha)^2 \bef )} \right)^{ k-\tilde k} \notag \\
&& \cdot \left ( \frac{\sqrt{8 (1 + 2 h^2\Pb)}}{27\sqrt{\pi} (1+ h^2 \Pb)} \right)^{\eta-k }\notag \\
&&  \cdot \left ( \frac{\sqrt{8 (1 + 2 h^2(1-\alpha)^2 \bef \Pb)}}{27\sqrt{\pi} (1+ h^2(1-\alpha)^2\bef \Pb)} \right )^{\tilde k}
\end{IEEEeqnarray}
\end{lemma}
\begin{IEEEproof}
similar to the proof of Lemmas~\ref{lemmaJU} and \ref{lemmaJe}. 
\end{IEEEproof}
As a result, we have 
\begin{IEEEeqnarray}{rCl}
\lefteqn{\Pr \Big [i^{(\e)}_{\text{SIC}} \Big( \{\xkes\}_{b \notin \bku  },  \{\xkef\}_{b \in \bku };} \notag \\
&& \hspace{0.5cm} \vect Y^{\Ne} | \bku , \bkut, \{\vect V_b\}_{b \in \bkut}\Big)  \le \tilde \gamma^{(\e)} \Big]  \\
&\le& \Pr \Big [\tilde i^{(\e)}_{\text{SIC}} \Big( \{\xkes \}_{b \notin \bku  },  \{\xkef\}_{b \in \bku }; \notag \\
&& \hspace{0.5cm} \vect Y^{\Ne} | \bku , \bkut,  \{\vect V_b\}_{b \in \bkut}\Big) < \tilde \gamma^{(\e)} - \ln \tilde J_{\e} \Big] \\
&=& \Pr \Bigg[ \ln \hspace{-0.15cm}\prod_{b\notin  \bku  }\hspace{-0.15cm} \frac{f_{\vect Y_b| \xkes} (\vect y_b| \vect x_{b}^{(\e,1)})}{Q_{\vect Y_b}(\vect y_b)} \notag \\
&& \hspace{0.2cm} +  \ln \hspace{-0.15cm}\prod_{b\in  \bku \backslash \bkut } \hspace{-0.15cm} \frac{Q_{\vect Y_b| \xkef} (\vect y_b| \vect x_{b}^{(\e,2)})}{Q_{\vect Y_b}(\vect y_b)}\notag \\
&& \hspace{0.2cm} +  \ln \hspace{-0.15cm}\prod_{b\in  \bkut } \hspace{-0.15cm} \frac{f_{\vect Y_b| \xkef, \vect V_b} (\vect y_b| \vect x_{b}^{(\e,2)}, \vect v_b)}{Q_{\vect Y_b|\vect V_b}(\vect y_b| \vect y_b)}< \tilde \gamma^{(\e)} - \ln \tilde J_{\e} \Bigg]  \IEEEeqnarraynumspace\\
&=&\Pr \Bigg[ \ln \hspace{-0.25cm}\prod_{b\notin  \bku\backslash \eta+1  }\hspace{-0.25cm} \frac{\frac{1}{(\sqrt{2\pi})^{\Nu}} e^{-\frac{||\vect Z_b||^2}{2}}}{\frac{1}{(\sqrt{2\pi \sy})^{\Nu}} e^{-\frac{||\xkes + \vect Z_b||^2}{2 \sy}}} \notag \\
&& \hspace{0.25cm} +  \ln \hspace{-0.15cm}\prod_{b\in  \bku \backslash \bkut  } \hspace{-0.15cm} \frac{\frac{1}{(\sqrt{2\pi \syx})^{\Nu}} e^{-\frac{||\vect V_b + \vect Z_b||^2}{2\syx}}}{\frac{1}{(\sqrt{2\pi \sy})^{\Nu}} e^{-\frac{||\xku + \xkef + \vect Z_b||^2}{2 \sy}}}\notag \\
&& \hspace{0.25cm} +  \ln \hspace{-0.15cm}\prod_{b\in  \bkut } \hspace{-0.15cm} \frac{\frac{1}{(\sqrt{2\pi})^{\Nu}} e^{-\frac{|| \vect Z_b||^2}{2}}}{\frac{1}{(\sqrt{2\pi \syv})^{\Nu}} e^{-\frac{||(1-\alpha) \xkef + \vect Z_b||^2}{2 \syv}}}\notag \\
&&  \hspace{0.25cm}+ \ln \frac{\frac{1}{(\sqrt{2\pi})^{\Ne-\eta\Nu}} e^{-\frac{|| \vect Z_{\eta+1}||^2}{2}}}{\frac{1}{(\sqrt{2\pi \sy})^{\Ne-\eta\Nu}} e^{-\frac{|| \vect X_{\eta+1,1}^{(\e)} + \vect Z_{\eta+1}||^2}{2 \sy}}}< \tilde \gamma^{(\e)} - \ln \tilde J_{\e} \Bigg]  \\
&=&\Pr \Bigg[  \frac{1}{2} \sum_{b \notin  \bku} ||\vect Z_b||^2 - \frac{1}{2\sy} ||\xkes + \vect Z_b||^2 \notag \\
&& \hspace{0.25cm}+  \sum_{b \in  \bku \backslash \bkut} \Bigg (\frac{||\vect V_b + \vect Z_b||^2}{2\syx} \notag \\
&& \hspace{1.7cm}  -  \frac{ ||\vect V_b + (1- \alpha)\xkef + \vect Z_b||^2}{2\sy} \Bigg) \notag \\
&& \hspace{0.25cm}+  \sum_{b \in  \bkut}\frac{|| \vect Z_b||^2}{2}  -  \frac{ || (1- \alpha)\xkef + \vect Z_b||^2}{2\syv} \notag \\
&&\hspace{0.25cm}> -\tilde \gamma^{(\e)} + \ln \tilde J_{\e}+\frac{\Ne - k \Nu}{2} \ln \sy  \notag \\
&& \hspace{0.5cm}+\frac{(k-\tilde k)\Nu}{2} \ln \frac{\sy}{\syx}+\frac{\Nu \tilde k}{2}  \ln \syv \Bigg] \\
&\stackrel{{(a)}}{\le}&  \Pr \Bigg[ \frac{\sy-1}{2\sy} \sum_{b \notin  \bku} ||\vect Z_b||^2 +   \frac{\sqrt{\Nu \Pb}}{\sy}\sum_{b \notin  \bku} ||\vect Z_b|| \notag \\
&&\hspace{0.25cm} +   \tau \sum_{b \in  \bku \backslash \bkut} ||\vect Z_b|| \notag \\
&& \hspace{1cm} +  \frac{\sy-\syx}{2\sy \syx}\sum_{b \in  \bku \backslash \bkut} ||\vect Z_b||^2 \notag \\
&&\hspace{1.5cm} +   \frac{(1-\alpha)\sqrt{\Nu \bef \Pb}}{\syv} \sum_{b \in  \bkut} ||\vect Z_b||  \notag \\
&&\hspace{2cm} +  \frac{\syv-1}{2\syv}\sum_{b \in  \bkut} ||\vect Z_b||^2  > \tilde \mu \Bigg]\\
& \le & \frac{\mathbb E \left [ \frac{\sy-1}{2\sy} \sum_{b \notin  \bku} ||\vect Z_b||^2 +   \frac{\sqrt{\Nu \Pb}}{\sy}\sum_{b \notin  \bku} ||\vect Z_b|| \right ]}{\tilde \mu} \notag \\
&& + \frac{\tau \mathbb E \left [\sum_{b \in  \bku \backslash \bkut} ||\vect Z_b|| \right ]}{\tilde \mu} \notag \\
&& + \frac{\mathbb E \left [ \frac{\sy-\syx}{2\sy \syx}\sum_{b \in  \bku \backslash \bkut} ||\vect Z_b||^2 \right ]}{\tilde \mu } \notag \\
&& + \frac{ \mathbb E \left [\frac{(1-\alpha)\sqrt{\Nu \bef \Pb}}{\syv} \sum_{b \in  \bkut} ||\vect Z_b|| \right ]}{\tilde \mu} \notag \\
&& + \frac{\mathbb E \left [ \frac{\syv-1}{2\syv}\sum_{b \in  \bkut} ||\vect Z_b||^2 \right ]}{ \tilde \mu} \notag \\
& = & \frac{(\Ne - k \Nu)(\sy-1)}{2\sy \tilde \mu} +  \frac{(\eta+1 - k)\sqrt{\Nu \Pb}}{\sy \tilde \mu} \frac{\sqrt{2} \Gamma\left (\frac{\Nu +1}{2}\right)}{\Gamma\left (\frac{\Nu}{2}\right)} \notag \\
&& + \frac{k\tau}{\tilde \mu}\frac{\sqrt{2} \Gamma\left (\frac{\Nu +1}{2}\right)}{\Gamma\left (\frac{\Nu}{2}\right)} +  \frac{k \Nu (\sy-\syx)}{2\sy \syx \tilde \mu} \notag \\
&& - \frac{\tilde k}{\tilde \mu}  \frac{\sqrt{2} \Gamma\left (\frac{\Nu +1}{2}\right)}{\Gamma\left (\frac{\Nu}{2}\right)}  \left (\tau -\frac{(1-\alpha)\sqrt{\Nu \bef \Pb}}{\syv}\right ) \notag \\
&& - \frac{\Nu \tilde k}{\tilde \mu} \left ( \frac{\sy-\syx}{2\sy \syx } - \frac{\syv-1}{2\syv}\right)
\end{IEEEeqnarray}
where 
\begin{IEEEeqnarray}{rCl}
\tilde \mu&: =&  \frac{\Ne - k\Nu}{2} \ln \sy  + \frac{(k - \tilde k)\Nu}{2} \ln \frac{\sy}{\syx}+ \frac{\tilde k\Nu}{2} \ln \syv \notag \\
&& - \frac{\eta-k}{2\sy} \Nu \Pb +   \frac{k- \tilde k}{2\syx}\vnorm \Nu \Pb  - \frac{\tilde k(1- \alpha)^2 \Nu \Pb \bef }{2 \syv}\notag \\
&& - \frac{k-\tilde k}{2\sy}\left (\sqrt{\vnorm} + (1- \alpha)\sqrt{\bef} \right)^2\Nu \Pb \notag  - \tilde \gamma^{(\e)} + \ln \tilde J_{\e}.
\end{IEEEeqnarray}
This concludes the proof.


\begin{thebibliography}{20}

\bibitem{Parkvall2017}
S.~Parkvall, E.~Dahlman, A.~Furuskar, and M.~Frenne, ``NR: The new 5G radio access technology," \emph{IEEE Communications Standards Magazine}, vol. 1, no. 4, pp. 24--30, Dec.~2017.

\bibitem{Abood2023}
M.~S.~Abood, H.~Wang, D.~He, Z.~Kang, and A.~Kawoya, ``Intelligent network slicing in V2X networks – A comprehensive review'',  \emph{Journal of Artificial Intelligence and Technology}, vol.~3,  no.~2, pp.~75--84, 2023. 

\bibitem{Noor2022}
M. Noor-A-Rahim et al., ``6G for vehicle-to-everything (V2X) communications: enabling technologies, challenges, and opportunities,"  \emph{Proceedings of the IEEE}, vol.~110, no.~6, pp. 712--734, June 2022.

\bibitem{Anand2020}
A.~Anand, G.~de Veciana, and S.~Shakkottai, ``Joint scheduling of URLLC and eMBB traffic in 5G wireless networks," \emph{IEEE/ACM Transactions on Networking}, vol. 28, no. 2, pp. 477-- 490, April 2020.


\bibitem{HomaEntropy2022}
H.~Nikbakht, M.~Wigger, M.~Egan, S.~Shamai (Shitz), J-M.~Gorce, and  H.~V.~Poor, ``An information-theoretic view of mixed-delay traffic in 5G and 6G," \emph{ Entropy}, vol.~24, no.~5, 2022. 

\bibitem{Popovski2018}
P.~Popovski, K.~F.~Trillingsgaard, O.~Simeone, and G. Durisi, ``5G wireless network slicing for eMBB, URLLC, and mMTC: A
communication-theoretic view,'' \emph{IEEE Access}, vol.~6, 2018.



\bibitem{YChen2020}
Y.~Chen, Y.~Wang, M.~Liu, J.~Zhang and L.~Jiao, ``Network slicing enabled resource management for service-oriented ultra-reliable and low-latency vehicular networks,"  \emph{IEEE Transactions on Vehicular Technology}, vol.~69, no. 7, pp.~7847--7862, July 2020.

\bibitem{Ganesan2019}
K.~Ganesan, P.~B.~Mallick, J.~Löhr, D.~Karampatsis, and A. Kunz, ``5G V2X architecture and radio aspects," in \emph{Proceeding of the IEEE Conference on Standards for Communications and Networking}, Granada, Spain, pp. 1--6, 2019.


\bibitem{Chen2020}
Q.~Chen, H.~Jiang, and G. Yu, ``Service oriented resource management in spatial reuse-based C-V2X networks," \emph{IEEE Wireless Communications Letters}, vol.~9, no.~1, pp.~91--94, Jan. ~2020. 

\bibitem{Yin2021}
H.~Yin, L.~Zhang, and S.~Roy, ``Multiplexing URLLC traffic within eMBB services in 5G NR: fair scheduling,"  \emph{IEEE Transactions on Communications}, vol.~69, no.~2, pp.~1080--1093, Feb. 2021.

\bibitem{Song2019}
X.~Song and M.~Yuan, ``Performance analysis of one-way highway vehicular networks with dynamic multiplexing of eMBB and URLLC traffics," \emph{IEEE Access}, vol.~7, pp. 118020--118029, 2019.

\bibitem{Costa1983}
M. H. M. Costa, ``Writing on dirty paper (Corresp.),'\emph{IEEE Transactions on Information Theory}, vol. 29, no. 3, pp. 439–441, May 1983.

  \bibitem{Scarlett2015}
  J. Scarlett, ``On the dispersions of the Gel'fand - Pinsker channel and dirty paper coding," \emph{IEEE Transactions on Information Theory}, vol. 61, no. 9, pp. 4569-4586, Sept. 2015.

\bibitem{Caire2003}
G. Caire and S. Shamai, ``On the achievable throughput of a multiantenna Gaussian broadcast channel,"  \emph{IEEE Transactions on Information Theory}, vol. 49, no. 7, pp. 1691-1706, July 2003.
\bibitem{Stam1982}
A.~J.~Stam, ``Limit theorems for uniform distributions on spheres in high-dimensional Euclidean spaces,” \emph{Journal of Applied Probability}, vol.~19, no.~1, pp. 221--228, 1982.

\bibitem{Molavianjazi}
E. MolavianJazi and J. N. Laneman, ``A second-order achievable rate region for Gaussian multi-access channels via a central limit theorem for functions,''  \emph{IEEE Transactions on Information Theory}, vol. 61, no. 12, pp. 6719--6733, Dec. 2015. 

\bibitem{Nikbakht2022}
H.~Nikbakht, M.~Wigger, S.~Shamai, J.~M.~Gorce, and H.~V.~Poor, ``Joint coding of URLLC and eMBB in Wyner's soft-handoff network in the finite blocklength regime," in \emph{Proceeding of the IEEE Global Communications Conference}, Rio de Janeiro, Brazil, pp.~1--6, 2022. 


 \end{thebibliography}
\end{document}